\documentclass{article}
\usepackage{color}
\usepackage{amsmath,amssymb}
\usepackage{amsbsy}
\usepackage{graphicx}
\usepackage{lineno}
\usepackage{comment}
\definecolor{AliceBlue}{rgb}{0.94,0.97,1.00}
\definecolor{AntiqueWhite1}{rgb}{1.00,0.94,0.86}
\definecolor{AntiqueWhite2}{rgb}{0.93,0.87,0.80}
\definecolor{AntiqueWhite3}{rgb}{0.80,0.75,0.69}
\definecolor{AntiqueWhite4}{rgb}{0.55,0.51,0.47}
\definecolor{AntiqueWhite}{rgb}{0.98,0.92,0.84}
\definecolor{BlanchedAlmond}{rgb}{1.00,0.92,0.80}
\definecolor{BlueViolet}{rgb}{0.54,0.17,0.89}
\definecolor{CadetBlue1}{rgb}{0.60,0.96,1.00}
\definecolor{CadetBlue2}{rgb}{0.56,0.90,0.93}
\definecolor{CadetBlue3}{rgb}{0.48,0.77,0.80}
\definecolor{CadetBlue4}{rgb}{0.33,0.53,0.55}
\definecolor{CadetBlue}{rgb}{0.37,0.62,0.63}
\definecolor{CornflowerBlue}{rgb}{0.39,0.58,0.93}
\definecolor{DarkBlue}{rgb}{0.00,0.00,0.55}
\definecolor{DarkCyan}{rgb}{0.00,0.55,0.55}
\definecolor{DarkGoldenrod1}{rgb}{1.00,0.73,0.06}
\definecolor{DarkGoldenrod2}{rgb}{0.93,0.68,0.05}
\definecolor{DarkGoldenrod3}{rgb}{0.80,0.58,0.05}
\definecolor{DarkGoldenrod4}{rgb}{0.55,0.40,0.03}
\definecolor{DarkGoldenrod}{rgb}{0.72,0.53,0.04}
\definecolor{DarkGray}{rgb}{0.66,0.66,0.66}
\definecolor{DarkGreen}{rgb}{0.00,0.39,0.00}
\definecolor{DarkGrey}{rgb}{0.66,0.66,0.66}
\definecolor{DarkKhaki}{rgb}{0.74,0.72,0.42}
\definecolor{DarkMagenta}{rgb}{0.55,0.00,0.55}
\definecolor{DarkOliveGreen1}{rgb}{0.79,1.00,0.44}
\definecolor{DarkOliveGreen2}{rgb}{0.74,0.93,0.41}
\definecolor{DarkOliveGreen3}{rgb}{0.64,0.80,0.35}
\definecolor{DarkOliveGreen4}{rgb}{0.43,0.55,0.24}
\definecolor{DarkOliveGreen}{rgb}{0.33,0.42,0.18}
\definecolor{DarkOrange1}{rgb}{1.00,0.50,0.00}
\definecolor{DarkOrange2}{rgb}{0.93,0.46,0.00}
\definecolor{DarkOrange3}{rgb}{0.80,0.40,0.00}
\definecolor{DarkOrange4}{rgb}{0.55,0.27,0.00}
\definecolor{DarkOrange}{rgb}{1.00,0.55,0.00}
\definecolor{DarkOrchid1}{rgb}{0.75,0.24,1.00}
\definecolor{DarkOrchid2}{rgb}{0.70,0.23,0.93}
\definecolor{DarkOrchid3}{rgb}{0.60,0.20,0.80}
\definecolor{DarkOrchid4}{rgb}{0.41,0.13,0.55}
\definecolor{DarkOrchid}{rgb}{0.60,0.20,0.80}
\definecolor{DarkRed}{rgb}{0.55,0.00,0.00}
\definecolor{DarkSalmon}{rgb}{0.91,0.59,0.48}
\definecolor{DarkSeaGreen1}{rgb}{0.76,1.00,0.76}
\definecolor{DarkSeaGreen2}{rgb}{0.71,0.93,0.71}
\definecolor{DarkSeaGreen3}{rgb}{0.61,0.80,0.61}
\definecolor{DarkSeaGreen4}{rgb}{0.41,0.55,0.41}
\definecolor{DarkSeaGreen}{rgb}{0.56,0.74,0.56}
\definecolor{DarkSlateBlue}{rgb}{0.28,0.24,0.55}
\definecolor{DarkSlateGray1}{rgb}{0.59,1.00,1.00}
\definecolor{DarkSlateGray2}{rgb}{0.55,0.93,0.93}
\definecolor{DarkSlateGray3}{rgb}{0.47,0.80,0.80}
\definecolor{DarkSlateGray4}{rgb}{0.32,0.55,0.55}
\definecolor{DarkSlateGray}{rgb}{0.18,0.31,0.31}
\definecolor{DarkSlateGrey}{rgb}{0.18,0.31,0.31}
\definecolor{DarkTurquoise}{rgb}{0.00,0.81,0.82}
\definecolor{DarkViolet}{rgb}{0.58,0.00,0.83}
\definecolor{DeepPink1}{rgb}{1.00,0.08,0.58}
\definecolor{DeepPink2}{rgb}{0.93,0.07,0.54}
\definecolor{DeepPink3}{rgb}{0.80,0.06,0.46}
\definecolor{DeepPink4}{rgb}{0.55,0.04,0.31}
\definecolor{DeepPink}{rgb}{1.00,0.08,0.58}
\definecolor{DeepSkyBlue1}{rgb}{0.00,0.75,1.00}
\definecolor{DeepSkyBlue2}{rgb}{0.00,0.70,0.93}
\definecolor{DeepSkyBlue3}{rgb}{0.00,0.60,0.80}
\definecolor{DeepSkyBlue4}{rgb}{0.00,0.41,0.55}
\definecolor{DeepSkyBlue}{rgb}{0.00,0.75,1.00}
\definecolor{DimGray}{rgb}{0.41,0.41,0.41}
\definecolor{DimGrey}{rgb}{0.41,0.41,0.41}
\definecolor{DodgerBlue1}{rgb}{0.12,0.56,1.00}
\definecolor{DodgerBlue2}{rgb}{0.11,0.53,0.93}
\definecolor{DodgerBlue3}{rgb}{0.09,0.45,0.80}
\definecolor{DodgerBlue4}{rgb}{0.06,0.31,0.55}
\definecolor{DodgerBlue}{rgb}{0.12,0.56,1.00}
\definecolor{FloralWhite}{rgb}{1.00,0.98,0.94}
\definecolor{ForestGreen}{rgb}{0.13,0.55,0.13}
\definecolor{GhostWhite}{rgb}{0.97,0.97,1.00}
\definecolor{GreenYellow}{rgb}{0.68,1.00,0.18}
\definecolor{HotPink1}{rgb}{1.00,0.43,0.71}
\definecolor{HotPink2}{rgb}{0.93,0.42,0.65}
\definecolor{HotPink3}{rgb}{0.80,0.38,0.56}
\definecolor{HotPink4}{rgb}{0.55,0.23,0.38}
\definecolor{HotPink}{rgb}{1.00,0.41,0.71}
\definecolor{IndianRed1}{rgb}{1.00,0.42,0.42}
\definecolor{IndianRed2}{rgb}{0.93,0.39,0.39}
\definecolor{IndianRed3}{rgb}{0.80,0.33,0.33}
\definecolor{IndianRed4}{rgb}{0.55,0.23,0.23}
\definecolor{IndianRed}{rgb}{0.80,0.36,0.36}
\definecolor{LavenderBlush1}{rgb}{1.00,0.94,0.96}
\definecolor{LavenderBlush2}{rgb}{0.93,0.88,0.90}
\definecolor{LavenderBlush3}{rgb}{0.80,0.76,0.77}
\definecolor{LavenderBlush4}{rgb}{0.55,0.51,0.53}
\definecolor{LavenderBlush}{rgb}{1.00,0.94,0.96}
\definecolor{LawnGreen}{rgb}{0.49,0.99,0.00}
\definecolor{LemonChiffon1}{rgb}{1.00,0.98,0.80}
\definecolor{LemonChiffon2}{rgb}{0.93,0.91,0.75}
\definecolor{LemonChiffon3}{rgb}{0.80,0.79,0.65}
\definecolor{LemonChiffon4}{rgb}{0.55,0.54,0.44}
\definecolor{LemonChiffon}{rgb}{1.00,0.98,0.80}
\definecolor{LightBlue1}{rgb}{0.75,0.94,1.00}
\definecolor{LightBlue2}{rgb}{0.70,0.87,0.93}
\definecolor{LightBlue3}{rgb}{0.60,0.75,0.80}
\definecolor{LightBlue4}{rgb}{0.41,0.51,0.55}
\definecolor{LightBlue}{rgb}{0.68,0.85,0.90}
\definecolor{LightCoral}{rgb}{0.94,0.50,0.50}
\definecolor{LightCyan1}{rgb}{0.88,1.00,1.00}
\definecolor{LightCyan2}{rgb}{0.82,0.93,0.93}
\definecolor{LightCyan3}{rgb}{0.71,0.80,0.80}
\definecolor{LightCyan4}{rgb}{0.48,0.55,0.55}
\definecolor{LightCyan}{rgb}{0.88,1.00,1.00}
\definecolor{LightGoldenrod1}{rgb}{1.00,0.93,0.55}
\definecolor{LightGoldenrod2}{rgb}{0.93,0.86,0.51}
\definecolor{LightGoldenrod3}{rgb}{0.80,0.75,0.44}
\definecolor{LightGoldenrod4}{rgb}{0.55,0.51,0.30}
\definecolor{LightGoldenrodYellow}{rgb}{0.98,0.98,0.82}
\definecolor{LightGoldenrod}{rgb}{0.93,0.87,0.51}
\definecolor{LightGray}{rgb}{0.83,0.83,0.83}
\definecolor{LightGreen}{rgb}{0.56,0.93,0.56}
\definecolor{LightGrey}{rgb}{0.83,0.83,0.83}
\definecolor{LightPink1}{rgb}{1.00,0.68,0.73}
\definecolor{LightPink2}{rgb}{0.93,0.64,0.68}
\definecolor{LightPink3}{rgb}{0.80,0.55,0.58}
\definecolor{LightPink4}{rgb}{0.55,0.37,0.40}
\definecolor{LightPink}{rgb}{1.00,0.71,0.76}
\definecolor{LightSalmon1}{rgb}{1.00,0.63,0.48}
\definecolor{LightSalmon2}{rgb}{0.93,0.58,0.45}
\definecolor{LightSalmon3}{rgb}{0.80,0.51,0.38}
\definecolor{LightSalmon4}{rgb}{0.55,0.34,0.26}
\definecolor{LightSalmon}{rgb}{1.00,0.63,0.48}
\definecolor{LightSeaGreen}{rgb}{0.13,0.70,0.67}
\definecolor{LightSkyBlue1}{rgb}{0.69,0.89,1.00}
\definecolor{LightSkyBlue2}{rgb}{0.64,0.83,0.93}
\definecolor{LightSkyBlue3}{rgb}{0.55,0.71,0.80}
\definecolor{LightSkyBlue4}{rgb}{0.38,0.48,0.55}
\definecolor{LightSkyBlue}{rgb}{0.53,0.81,0.98}
\definecolor{LightSlateBlue}{rgb}{0.52,0.44,1.00}
\definecolor{LightSlateGray}{rgb}{0.47,0.53,0.60}
\definecolor{LightSlateGrey}{rgb}{0.47,0.53,0.60}
\definecolor{LightSteelBlue1}{rgb}{0.79,0.88,1.00}
\definecolor{LightSteelBlue2}{rgb}{0.74,0.82,0.93}
\definecolor{LightSteelBlue3}{rgb}{0.64,0.71,0.80}
\definecolor{LightSteelBlue4}{rgb}{0.43,0.48,0.55}
\definecolor{LightSteelBlue}{rgb}{0.69,0.77,0.87}
\definecolor{LightYellow1}{rgb}{1.00,1.00,0.88}
\definecolor{LightYellow2}{rgb}{0.93,0.93,0.82}
\definecolor{LightYellow3}{rgb}{0.80,0.80,0.71}
\definecolor{LightYellow4}{rgb}{0.55,0.55,0.48}
\definecolor{LightYellow}{rgb}{1.00,1.00,0.88}
\definecolor{LimeGreen}{rgb}{0.20,0.80,0.20}
\definecolor{MediumAquamarine}{rgb}{0.40,0.80,0.67}
\definecolor{MediumBlue}{rgb}{0.00,0.00,0.80}
\definecolor{MediumOrchid1}{rgb}{0.88,0.40,1.00}
\definecolor{MediumOrchid2}{rgb}{0.82,0.37,0.93}
\definecolor{MediumOrchid3}{rgb}{0.71,0.32,0.80}
\definecolor{MediumOrchid4}{rgb}{0.48,0.22,0.55}
\definecolor{MediumOrchid}{rgb}{0.73,0.33,0.83}
\definecolor{MediumPurple1}{rgb}{0.67,0.51,1.00}
\definecolor{MediumPurple2}{rgb}{0.62,0.47,0.93}
\definecolor{MediumPurple3}{rgb}{0.54,0.41,0.80}
\definecolor{MediumPurple4}{rgb}{0.36,0.28,0.55}
\definecolor{MediumPurple}{rgb}{0.58,0.44,0.86}
\definecolor{MediumSeaGreen}{rgb}{0.24,0.70,0.44}
\definecolor{MediumSlateBlue}{rgb}{0.48,0.41,0.93}
\definecolor{MediumSpringGreen}{rgb}{0.00,0.98,0.60}
\definecolor{MediumTurquoise}{rgb}{0.28,0.82,0.80}
\definecolor{MediumVioletRed}{rgb}{0.78,0.08,0.52}
\definecolor{MidnightBlue}{rgb}{0.10,0.10,0.44}
\definecolor{MintCream}{rgb}{0.96,1.00,0.98}
\definecolor{MistyRose1}{rgb}{1.00,0.89,0.88}
\definecolor{MistyRose2}{rgb}{0.93,0.84,0.82}
\definecolor{MistyRose3}{rgb}{0.80,0.72,0.71}
\definecolor{MistyRose4}{rgb}{0.55,0.49,0.48}
\definecolor{MistyRose}{rgb}{1.00,0.89,0.88}
\definecolor{NavajoWhite1}{rgb}{1.00,0.87,0.68}
\definecolor{NavajoWhite2}{rgb}{0.93,0.81,0.63}
\definecolor{NavajoWhite3}{rgb}{0.80,0.70,0.55}
\definecolor{NavajoWhite4}{rgb}{0.55,0.47,0.37}
\definecolor{NavajoWhite}{rgb}{1.00,0.87,0.68}
\definecolor{NavyBlue}{rgb}{0.00,0.00,0.50}
\definecolor{OldLace}{rgb}{0.99,0.96,0.90}
\definecolor{OliveDrab1}{rgb}{0.75,1.00,0.24}
\definecolor{OliveDrab2}{rgb}{0.70,0.93,0.23}
\definecolor{OliveDrab3}{rgb}{0.60,0.80,0.20}
\definecolor{OliveDrab4}{rgb}{0.41,0.55,0.13}
\definecolor{OliveDrab}{rgb}{0.42,0.56,0.14}
\definecolor{OrangeRed1}{rgb}{1.00,0.27,0.00}
\definecolor{OrangeRed2}{rgb}{0.93,0.25,0.00}
\definecolor{OrangeRed3}{rgb}{0.80,0.22,0.00}
\definecolor{OrangeRed4}{rgb}{0.55,0.15,0.00}
\definecolor{OrangeRed}{rgb}{1.00,0.27,0.00}
\definecolor{PaleGoldenrod}{rgb}{0.93,0.91,0.67}
\definecolor{PaleGreen1}{rgb}{0.60,1.00,0.60}
\definecolor{PaleGreen2}{rgb}{0.56,0.93,0.56}
\definecolor{PaleGreen3}{rgb}{0.49,0.80,0.49}
\definecolor{PaleGreen4}{rgb}{0.33,0.55,0.33}
\definecolor{PaleGreen}{rgb}{0.60,0.98,0.60}
\definecolor{PaleTurquoise1}{rgb}{0.73,1.00,1.00}
\definecolor{PaleTurquoise2}{rgb}{0.68,0.93,0.93}
\definecolor{PaleTurquoise3}{rgb}{0.59,0.80,0.80}
\definecolor{PaleTurquoise4}{rgb}{0.40,0.55,0.55}
\definecolor{PaleTurquoise}{rgb}{0.69,0.93,0.93}
\definecolor{PaleVioletRed1}{rgb}{1.00,0.51,0.67}
\definecolor{PaleVioletRed2}{rgb}{0.93,0.47,0.62}
\definecolor{PaleVioletRed3}{rgb}{0.80,0.41,0.54}
\definecolor{PaleVioletRed4}{rgb}{0.55,0.28,0.36}
\definecolor{PaleVioletRed}{rgb}{0.86,0.44,0.58}
\definecolor{PapayaWhip}{rgb}{1.00,0.94,0.84}
\definecolor{PeachPuff1}{rgb}{1.00,0.85,0.73}
\definecolor{PeachPuff2}{rgb}{0.93,0.80,0.68}
\definecolor{PeachPuff3}{rgb}{0.80,0.69,0.58}
\definecolor{PeachPuff4}{rgb}{0.55,0.47,0.40}
\definecolor{PeachPuff}{rgb}{1.00,0.85,0.73}
\definecolor{PowderBlue}{rgb}{0.69,0.88,0.90}
\definecolor{RosyBrown1}{rgb}{1.00,0.76,0.76}
\definecolor{RosyBrown2}{rgb}{0.93,0.71,0.71}
\definecolor{RosyBrown3}{rgb}{0.80,0.61,0.61}
\definecolor{RosyBrown4}{rgb}{0.55,0.41,0.41}
\definecolor{RosyBrown}{rgb}{0.74,0.56,0.56}
\definecolor{RoyalBlue1}{rgb}{0.28,0.46,1.00}
\definecolor{RoyalBlue2}{rgb}{0.26,0.43,0.93}
\definecolor{RoyalBlue3}{rgb}{0.23,0.37,0.80}
\definecolor{RoyalBlue4}{rgb}{0.15,0.25,0.55}
\definecolor{RoyalBlue}{rgb}{0.25,0.41,0.88}
\definecolor{SaddleBrown}{rgb}{0.55,0.27,0.07}
\definecolor{SandyBrown}{rgb}{0.96,0.64,0.38}
\definecolor{SeaGreen1}{rgb}{0.33,1.00,0.62}
\definecolor{SeaGreen2}{rgb}{0.31,0.93,0.58}
\definecolor{SeaGreen3}{rgb}{0.26,0.80,0.50}
\definecolor{SeaGreen4}{rgb}{0.18,0.55,0.34}
\definecolor{SeaGreen}{rgb}{0.18,0.55,0.34}
\definecolor{SkyBlue1}{rgb}{0.53,0.81,1.00}
\definecolor{SkyBlue2}{rgb}{0.49,0.75,0.93}
\definecolor{SkyBlue3}{rgb}{0.42,0.65,0.80}
\definecolor{SkyBlue4}{rgb}{0.29,0.44,0.55}
\definecolor{SkyBlue}{rgb}{0.53,0.81,0.92}
\definecolor{SlateBlue1}{rgb}{0.51,0.44,1.00}
\definecolor{SlateBlue2}{rgb}{0.48,0.40,0.93}
\definecolor{SlateBlue3}{rgb}{0.41,0.35,0.80}
\definecolor{SlateBlue4}{rgb}{0.28,0.24,0.55}
\definecolor{SlateBlue}{rgb}{0.42,0.35,0.80}
\definecolor{SlateGray1}{rgb}{0.78,0.89,1.00}
\definecolor{SlateGray2}{rgb}{0.73,0.83,0.93}
\definecolor{SlateGray3}{rgb}{0.62,0.71,0.80}
\definecolor{SlateGray4}{rgb}{0.42,0.48,0.55}
\definecolor{SlateGray}{rgb}{0.44,0.50,0.56}
\definecolor{SlateGrey}{rgb}{0.44,0.50,0.56}
\definecolor{SpringGreen1}{rgb}{0.00,1.00,0.50}
\definecolor{SpringGreen2}{rgb}{0.00,0.93,0.46}
\definecolor{SpringGreen3}{rgb}{0.00,0.80,0.40}
\definecolor{SpringGreen4}{rgb}{0.00,0.55,0.27}
\definecolor{SpringGreen}{rgb}{0.00,1.00,0.50}
\definecolor{SteelBlue1}{rgb}{0.39,0.72,1.00}
\definecolor{SteelBlue2}{rgb}{0.36,0.67,0.93}
\definecolor{SteelBlue3}{rgb}{0.31,0.58,0.80}
\definecolor{SteelBlue4}{rgb}{0.21,0.39,0.55}
\definecolor{SteelBlue}{rgb}{0.27,0.51,0.71}
\definecolor{VioletRed1}{rgb}{1.00,0.24,0.59}
\definecolor{VioletRed2}{rgb}{0.93,0.23,0.55}
\definecolor{VioletRed3}{rgb}{0.80,0.20,0.47}
\definecolor{VioletRed4}{rgb}{0.55,0.13,0.32}
\definecolor{VioletRed}{rgb}{0.82,0.13,0.56}
\definecolor{WhiteSmoke}{rgb}{0.96,0.96,0.96}
\definecolor{YellowGreen}{rgb}{0.60,0.80,0.20}
\definecolor{aliceblue}{rgb}{0.94,0.97,1.00}
\definecolor{antiquewhite}{rgb}{0.98,0.92,0.84}
\definecolor{aquamarine1}{rgb}{0.50,1.00,0.83}
\definecolor{aquamarine2}{rgb}{0.46,0.93,0.78}
\definecolor{aquamarine3}{rgb}{0.40,0.80,0.67}
\definecolor{aquamarine4}{rgb}{0.27,0.55,0.45}
\definecolor{aquamarine}{rgb}{0.50,1.00,0.83}
\definecolor{azure1}{rgb}{0.94,1.00,1.00}
\definecolor{azure2}{rgb}{0.88,0.93,0.93}
\definecolor{azure3}{rgb}{0.76,0.80,0.80}
\definecolor{azure4}{rgb}{0.51,0.55,0.55}
\definecolor{azure}{rgb}{0.94,1.00,1.00}
\definecolor{beige}{rgb}{0.96,0.96,0.86}
\definecolor{bisque1}{rgb}{1.00,0.89,0.77}
\definecolor{bisque2}{rgb}{0.93,0.84,0.72}
\definecolor{bisque3}{rgb}{0.80,0.72,0.62}
\definecolor{bisque4}{rgb}{0.55,0.49,0.42}
\definecolor{bisque}{rgb}{1.00,0.89,0.77}
\definecolor{black}{rgb}{0.00,0.00,0.00}
\definecolor{blanchedalmond}{rgb}{1.00,0.92,0.80}
\definecolor{blue1}{rgb}{0.00,0.00,1.00}
\definecolor{blue2}{rgb}{0.00,0.00,0.93}
\definecolor{blue3}{rgb}{0.00,0.00,0.80}
\definecolor{blue4}{rgb}{0.00,0.00,0.55}
\definecolor{blueviolet}{rgb}{0.54,0.17,0.89}
\definecolor{blue}{rgb}{0.00,0.00,1.00}
\definecolor{brown1}{rgb}{1.00,0.25,0.25}
\definecolor{brown2}{rgb}{0.93,0.23,0.23}
\definecolor{brown3}{rgb}{0.80,0.20,0.20}
\definecolor{brown4}{rgb}{0.55,0.14,0.14}
\definecolor{brown}{rgb}{0.65,0.16,0.16}
\definecolor{burlywood1}{rgb}{1.00,0.83,0.61}
\definecolor{burlywood2}{rgb}{0.93,0.77,0.57}
\definecolor{burlywood3}{rgb}{0.80,0.67,0.49}
\definecolor{burlywood4}{rgb}{0.55,0.45,0.33}
\definecolor{burlywood}{rgb}{0.87,0.72,0.53}
\definecolor{cadetblue}{rgb}{0.37,0.62,0.63}
\definecolor{chartreuse1}{rgb}{0.50,1.00,0.00}
\definecolor{chartreuse2}{rgb}{0.46,0.93,0.00}
\definecolor{chartreuse3}{rgb}{0.40,0.80,0.00}
\definecolor{chartreuse4}{rgb}{0.27,0.55,0.00}
\definecolor{chartreuse}{rgb}{0.50,1.00,0.00}
\definecolor{chocolate1}{rgb}{1.00,0.50,0.14}
\definecolor{chocolate2}{rgb}{0.93,0.46,0.13}
\definecolor{chocolate3}{rgb}{0.80,0.40,0.11}
\definecolor{chocolate4}{rgb}{0.55,0.27,0.07}
\definecolor{chocolate}{rgb}{0.82,0.41,0.12}
\definecolor{coral1}{rgb}{1.00,0.45,0.34}
\definecolor{coral2}{rgb}{0.93,0.42,0.31}
\definecolor{coral3}{rgb}{0.80,0.36,0.27}
\definecolor{coral4}{rgb}{0.55,0.24,0.18}
\definecolor{coral}{rgb}{1.00,0.50,0.31}
\definecolor{cornflowerblue}{rgb}{0.39,0.58,0.93}
\definecolor{cornsilk1}{rgb}{1.00,0.97,0.86}
\definecolor{cornsilk2}{rgb}{0.93,0.91,0.80}
\definecolor{cornsilk3}{rgb}{0.80,0.78,0.69}
\definecolor{cornsilk4}{rgb}{0.55,0.53,0.47}
\definecolor{cornsilk}{rgb}{1.00,0.97,0.86}
\definecolor{cyan1}{rgb}{0.00,1.00,1.00}
\definecolor{cyan2}{rgb}{0.00,0.93,0.93}
\definecolor{cyan3}{rgb}{0.00,0.80,0.80}
\definecolor{cyan4}{rgb}{0.00,0.55,0.55}
\definecolor{cyan}{rgb}{0.00,1.00,1.00}
\definecolor{darkblue}{rgb}{0.00,0.00,0.55}
\definecolor{darkcyan}{rgb}{0.00,0.55,0.55}
\definecolor{darkgoldenrod}{rgb}{0.72,0.53,0.04}
\definecolor{darkgray}{rgb}{0.66,0.66,0.66}
\definecolor{darkgreen}{rgb}{0.00,0.39,0.00}
\definecolor{darkgrey}{rgb}{0.66,0.66,0.66}
\definecolor{darkkhaki}{rgb}{0.74,0.72,0.42}
\definecolor{darkmagenta}{rgb}{0.55,0.00,0.55}
\definecolor{darkolive}{rgb}{0.33,0.42,0.18}
\definecolor{darkorange}{rgb}{1.00,0.55,0.00}
\definecolor{darkorchid}{rgb}{0.60,0.20,0.80}
\definecolor{darkred}{rgb}{0.55,0.00,0.00}
\definecolor{darksalmon}{rgb}{0.91,0.59,0.48}
\definecolor{darksea}{rgb}{0.56,0.74,0.56}
\definecolor{darkslate}{rgb}{0.18,0.31,0.31}
\definecolor{darkslate}{rgb}{0.18,0.31,0.31}
\definecolor{darkslate}{rgb}{0.28,0.24,0.55}
\definecolor{darkturquoise}{rgb}{0.00,0.81,0.82}
\definecolor{darkviolet}{rgb}{0.58,0.00,0.83}
\definecolor{deeppink}{rgb}{1.00,0.08,0.58}
\definecolor{deepsky}{rgb}{0.00,0.75,1.00}
\definecolor{dimgray}{rgb}{0.41,0.41,0.41}
\definecolor{dimgrey}{rgb}{0.41,0.41,0.41}
\definecolor{dodgerblue}{rgb}{0.12,0.56,1.00}
\definecolor{firebrick1}{rgb}{1.00,0.19,0.19}
\definecolor{firebrick2}{rgb}{0.93,0.17,0.17}
\definecolor{firebrick3}{rgb}{0.80,0.15,0.15}
\definecolor{firebrick4}{rgb}{0.55,0.10,0.10}
\definecolor{firebrick}{rgb}{0.70,0.13,0.13}
\definecolor{floralwhite}{rgb}{1.00,0.98,0.94}
\definecolor{forestgreen}{rgb}{0.13,0.55,0.13}
\definecolor{gainsboro}{rgb}{0.86,0.86,0.86}
\definecolor{ghostwhite}{rgb}{0.97,0.97,1.00}
\definecolor{gold1}{rgb}{1.00,0.84,0.00}
\definecolor{gold2}{rgb}{0.93,0.79,0.00}
\definecolor{gold3}{rgb}{0.80,0.68,0.00}
\definecolor{gold4}{rgb}{0.55,0.46,0.00}
\definecolor{goldenrod1}{rgb}{1.00,0.76,0.15}
\definecolor{goldenrod2}{rgb}{0.93,0.71,0.13}
\definecolor{goldenrod3}{rgb}{0.80,0.61,0.11}
\definecolor{goldenrod4}{rgb}{0.55,0.41,0.08}
\definecolor{goldenrod}{rgb}{0.85,0.65,0.13}
\definecolor{gold}{rgb}{1.00,0.84,0.00}
\definecolor{gray0}{rgb}{0.00,0.00,0.00}
\definecolor{gray100}{rgb}{1.00,1.00,1.00}
\definecolor{gray10}{rgb}{0.10,0.10,0.10}
\definecolor{gray11}{rgb}{0.11,0.11,0.11}
\definecolor{gray12}{rgb}{0.12,0.12,0.12}
\definecolor{gray13}{rgb}{0.13,0.13,0.13}
\definecolor{gray14}{rgb}{0.14,0.14,0.14}
\definecolor{gray15}{rgb}{0.15,0.15,0.15}
\definecolor{gray16}{rgb}{0.16,0.16,0.16}
\definecolor{gray17}{rgb}{0.17,0.17,0.17}
\definecolor{gray18}{rgb}{0.18,0.18,0.18}
\definecolor{gray19}{rgb}{0.19,0.19,0.19}
\definecolor{gray1}{rgb}{0.01,0.01,0.01}
\definecolor{gray20}{rgb}{0.20,0.20,0.20}
\definecolor{gray21}{rgb}{0.21,0.21,0.21}
\definecolor{gray22}{rgb}{0.22,0.22,0.22}
\definecolor{gray23}{rgb}{0.23,0.23,0.23}
\definecolor{gray24}{rgb}{0.24,0.24,0.24}
\definecolor{gray25}{rgb}{0.25,0.25,0.25}
\definecolor{gray26}{rgb}{0.26,0.26,0.26}
\definecolor{gray27}{rgb}{0.27,0.27,0.27}
\definecolor{gray28}{rgb}{0.28,0.28,0.28}
\definecolor{gray29}{rgb}{0.29,0.29,0.29}
\definecolor{gray2}{rgb}{0.02,0.02,0.02}
\definecolor{gray30}{rgb}{0.30,0.30,0.30}
\definecolor{gray31}{rgb}{0.31,0.31,0.31}
\definecolor{gray32}{rgb}{0.32,0.32,0.32}
\definecolor{gray33}{rgb}{0.33,0.33,0.33}
\definecolor{gray34}{rgb}{0.34,0.34,0.34}
\definecolor{gray35}{rgb}{0.35,0.35,0.35}
\definecolor{gray36}{rgb}{0.36,0.36,0.36}
\definecolor{gray37}{rgb}{0.37,0.37,0.37}
\definecolor{gray38}{rgb}{0.38,0.38,0.38}
\definecolor{gray39}{rgb}{0.39,0.39,0.39}
\definecolor{gray3}{rgb}{0.03,0.03,0.03}
\definecolor{gray40}{rgb}{0.40,0.40,0.40}
\definecolor{gray41}{rgb}{0.41,0.41,0.41}
\definecolor{gray42}{rgb}{0.42,0.42,0.42}
\definecolor{gray43}{rgb}{0.43,0.43,0.43}
\definecolor{gray44}{rgb}{0.44,0.44,0.44}
\definecolor{gray45}{rgb}{0.45,0.45,0.45}
\definecolor{gray46}{rgb}{0.46,0.46,0.46}
\definecolor{gray47}{rgb}{0.47,0.47,0.47}
\definecolor{gray48}{rgb}{0.48,0.48,0.48}
\definecolor{gray49}{rgb}{0.49,0.49,0.49}
\definecolor{gray4}{rgb}{0.04,0.04,0.04}
\definecolor{gray50}{rgb}{0.50,0.50,0.50}
\definecolor{gray51}{rgb}{0.51,0.51,0.51}
\definecolor{gray52}{rgb}{0.52,0.52,0.52}
\definecolor{gray53}{rgb}{0.53,0.53,0.53}
\definecolor{gray54}{rgb}{0.54,0.54,0.54}
\definecolor{gray55}{rgb}{0.55,0.55,0.55}
\definecolor{gray56}{rgb}{0.56,0.56,0.56}
\definecolor{gray57}{rgb}{0.57,0.57,0.57}
\definecolor{gray58}{rgb}{0.58,0.58,0.58}
\definecolor{gray59}{rgb}{0.59,0.59,0.59}
\definecolor{gray5}{rgb}{0.05,0.05,0.05}
\definecolor{gray60}{rgb}{0.60,0.60,0.60}
\definecolor{gray61}{rgb}{0.61,0.61,0.61}
\definecolor{gray62}{rgb}{0.62,0.62,0.62}
\definecolor{gray63}{rgb}{0.63,0.63,0.63}
\definecolor{gray64}{rgb}{0.64,0.64,0.64}
\definecolor{gray65}{rgb}{0.65,0.65,0.65}
\definecolor{gray66}{rgb}{0.66,0.66,0.66}
\definecolor{gray67}{rgb}{0.67,0.67,0.67}
\definecolor{gray68}{rgb}{0.68,0.68,0.68}
\definecolor{gray69}{rgb}{0.69,0.69,0.69}
\definecolor{gray6}{rgb}{0.06,0.06,0.06}
\definecolor{gray70}{rgb}{0.70,0.70,0.70}
\definecolor{gray71}{rgb}{0.71,0.71,0.71}
\definecolor{gray72}{rgb}{0.72,0.72,0.72}
\definecolor{gray73}{rgb}{0.73,0.73,0.73}
\definecolor{gray74}{rgb}{0.74,0.74,0.74}
\definecolor{gray75}{rgb}{0.75,0.75,0.75}
\definecolor{gray76}{rgb}{0.76,0.76,0.76}
\definecolor{gray77}{rgb}{0.77,0.77,0.77}
\definecolor{gray78}{rgb}{0.78,0.78,0.78}
\definecolor{gray79}{rgb}{0.79,0.79,0.79}
\definecolor{gray7}{rgb}{0.07,0.07,0.07}
\definecolor{gray80}{rgb}{0.80,0.80,0.80}
\definecolor{gray81}{rgb}{0.81,0.81,0.81}
\definecolor{gray82}{rgb}{0.82,0.82,0.82}
\definecolor{gray83}{rgb}{0.83,0.83,0.83}
\definecolor{gray84}{rgb}{0.84,0.84,0.84}
\definecolor{gray85}{rgb}{0.85,0.85,0.85}
\definecolor{gray86}{rgb}{0.86,0.86,0.86}
\definecolor{gray87}{rgb}{0.87,0.87,0.87}
\definecolor{gray88}{rgb}{0.88,0.88,0.88}
\definecolor{gray89}{rgb}{0.89,0.89,0.89}
\definecolor{gray8}{rgb}{0.08,0.08,0.08}
\definecolor{gray90}{rgb}{0.90,0.90,0.90}
\definecolor{gray91}{rgb}{0.91,0.91,0.91}
\definecolor{gray92}{rgb}{0.92,0.92,0.92}
\definecolor{gray93}{rgb}{0.93,0.93,0.93}
\definecolor{gray94}{rgb}{0.94,0.94,0.94}
\definecolor{gray95}{rgb}{0.95,0.95,0.95}
\definecolor{gray96}{rgb}{0.96,0.96,0.96}
\definecolor{gray97}{rgb}{0.97,0.97,0.97}
\definecolor{gray98}{rgb}{0.98,0.98,0.98}
\definecolor{gray99}{rgb}{0.99,0.99,0.99}
\definecolor{gray9}{rgb}{0.09,0.09,0.09}
\definecolor{gray}{rgb}{0.75,0.75,0.75}
\definecolor{green1}{rgb}{0.00,1.00,0.00}
\definecolor{green2}{rgb}{0.00,0.93,0.00}
\definecolor{green3}{rgb}{0.00,0.80,0.00}
\definecolor{green4}{rgb}{0.00,0.55,0.00}
\definecolor{greenyellow}{rgb}{0.68,1.00,0.18}
\definecolor{green}{rgb}{0.00,1.00,0.00}
\definecolor{grey0}{rgb}{0.00,0.00,0.00}
\definecolor{grey100}{rgb}{1.00,1.00,1.00}
\definecolor{grey10}{rgb}{0.10,0.10,0.10}
\definecolor{grey11}{rgb}{0.11,0.11,0.11}
\definecolor{grey12}{rgb}{0.12,0.12,0.12}
\definecolor{grey13}{rgb}{0.13,0.13,0.13}
\definecolor{grey14}{rgb}{0.14,0.14,0.14}
\definecolor{grey15}{rgb}{0.15,0.15,0.15}
\definecolor{grey16}{rgb}{0.16,0.16,0.16}
\definecolor{grey17}{rgb}{0.17,0.17,0.17}
\definecolor{grey18}{rgb}{0.18,0.18,0.18}
\definecolor{grey19}{rgb}{0.19,0.19,0.19}
\definecolor{grey1}{rgb}{0.01,0.01,0.01}
\definecolor{grey20}{rgb}{0.20,0.20,0.20}
\definecolor{grey21}{rgb}{0.21,0.21,0.21}
\definecolor{grey22}{rgb}{0.22,0.22,0.22}
\definecolor{grey23}{rgb}{0.23,0.23,0.23}
\definecolor{grey24}{rgb}{0.24,0.24,0.24}
\definecolor{grey25}{rgb}{0.25,0.25,0.25}
\definecolor{grey26}{rgb}{0.26,0.26,0.26}
\definecolor{grey27}{rgb}{0.27,0.27,0.27}
\definecolor{grey28}{rgb}{0.28,0.28,0.28}
\definecolor{grey29}{rgb}{0.29,0.29,0.29}
\definecolor{grey2}{rgb}{0.02,0.02,0.02}
\definecolor{grey30}{rgb}{0.30,0.30,0.30}
\definecolor{grey31}{rgb}{0.31,0.31,0.31}
\definecolor{grey32}{rgb}{0.32,0.32,0.32}
\definecolor{grey33}{rgb}{0.33,0.33,0.33}
\definecolor{grey34}{rgb}{0.34,0.34,0.34}
\definecolor{grey35}{rgb}{0.35,0.35,0.35}
\definecolor{grey36}{rgb}{0.36,0.36,0.36}
\definecolor{grey37}{rgb}{0.37,0.37,0.37}
\definecolor{grey38}{rgb}{0.38,0.38,0.38}
\definecolor{grey39}{rgb}{0.39,0.39,0.39}
\definecolor{grey3}{rgb}{0.03,0.03,0.03}
\definecolor{grey40}{rgb}{0.40,0.40,0.40}
\definecolor{grey41}{rgb}{0.41,0.41,0.41}
\definecolor{grey42}{rgb}{0.42,0.42,0.42}
\definecolor{grey43}{rgb}{0.43,0.43,0.43}
\definecolor{grey44}{rgb}{0.44,0.44,0.44}
\definecolor{grey45}{rgb}{0.45,0.45,0.45}
\definecolor{grey46}{rgb}{0.46,0.46,0.46}
\definecolor{grey47}{rgb}{0.47,0.47,0.47}
\definecolor{grey48}{rgb}{0.48,0.48,0.48}
\definecolor{grey49}{rgb}{0.49,0.49,0.49}
\definecolor{grey4}{rgb}{0.04,0.04,0.04}
\definecolor{grey50}{rgb}{0.50,0.50,0.50}
\definecolor{grey51}{rgb}{0.51,0.51,0.51}
\definecolor{grey52}{rgb}{0.52,0.52,0.52}
\definecolor{grey53}{rgb}{0.53,0.53,0.53}
\definecolor{grey54}{rgb}{0.54,0.54,0.54}
\definecolor{grey55}{rgb}{0.55,0.55,0.55}
\definecolor{grey56}{rgb}{0.56,0.56,0.56}
\definecolor{grey57}{rgb}{0.57,0.57,0.57}
\definecolor{grey58}{rgb}{0.58,0.58,0.58}
\definecolor{grey59}{rgb}{0.59,0.59,0.59}
\definecolor{grey5}{rgb}{0.05,0.05,0.05}
\definecolor{grey60}{rgb}{0.60,0.60,0.60}
\definecolor{grey61}{rgb}{0.61,0.61,0.61}
\definecolor{grey62}{rgb}{0.62,0.62,0.62}
\definecolor{grey63}{rgb}{0.63,0.63,0.63}
\definecolor{grey64}{rgb}{0.64,0.64,0.64}
\definecolor{grey65}{rgb}{0.65,0.65,0.65}
\definecolor{grey66}{rgb}{0.66,0.66,0.66}
\definecolor{grey67}{rgb}{0.67,0.67,0.67}
\definecolor{grey68}{rgb}{0.68,0.68,0.68}
\definecolor{grey69}{rgb}{0.69,0.69,0.69}
\definecolor{grey6}{rgb}{0.06,0.06,0.06}
\definecolor{grey70}{rgb}{0.70,0.70,0.70}
\definecolor{grey71}{rgb}{0.71,0.71,0.71}
\definecolor{grey72}{rgb}{0.72,0.72,0.72}
\definecolor{grey73}{rgb}{0.73,0.73,0.73}
\definecolor{grey74}{rgb}{0.74,0.74,0.74}
\definecolor{grey75}{rgb}{0.75,0.75,0.75}
\definecolor{grey76}{rgb}{0.76,0.76,0.76}
\definecolor{grey77}{rgb}{0.77,0.77,0.77}
\definecolor{grey78}{rgb}{0.78,0.78,0.78}
\definecolor{grey79}{rgb}{0.79,0.79,0.79}
\definecolor{grey7}{rgb}{0.07,0.07,0.07}
\definecolor{grey80}{rgb}{0.80,0.80,0.80}
\definecolor{grey81}{rgb}{0.81,0.81,0.81}
\definecolor{grey82}{rgb}{0.82,0.82,0.82}
\definecolor{grey83}{rgb}{0.83,0.83,0.83}
\definecolor{grey84}{rgb}{0.84,0.84,0.84}
\definecolor{grey85}{rgb}{0.85,0.85,0.85}
\definecolor{grey86}{rgb}{0.86,0.86,0.86}
\definecolor{grey87}{rgb}{0.87,0.87,0.87}
\definecolor{grey88}{rgb}{0.88,0.88,0.88}
\definecolor{grey89}{rgb}{0.89,0.89,0.89}
\definecolor{grey8}{rgb}{0.08,0.08,0.08}
\definecolor{grey90}{rgb}{0.90,0.90,0.90}
\definecolor{grey91}{rgb}{0.91,0.91,0.91}
\definecolor{grey92}{rgb}{0.92,0.92,0.92}
\definecolor{grey93}{rgb}{0.93,0.93,0.93}
\definecolor{grey94}{rgb}{0.94,0.94,0.94}
\definecolor{grey95}{rgb}{0.95,0.95,0.95}
\definecolor{grey96}{rgb}{0.96,0.96,0.96}
\definecolor{grey97}{rgb}{0.97,0.97,0.97}
\definecolor{grey98}{rgb}{0.98,0.98,0.98}
\definecolor{grey99}{rgb}{0.99,0.99,0.99}
\definecolor{grey9}{rgb}{0.09,0.09,0.09}
\definecolor{grey}{rgb}{0.75,0.75,0.75}
\definecolor{honeydew1}{rgb}{0.94,1.00,0.94}
\definecolor{honeydew2}{rgb}{0.88,0.93,0.88}
\definecolor{honeydew3}{rgb}{0.76,0.80,0.76}
\definecolor{honeydew4}{rgb}{0.51,0.55,0.51}
\definecolor{honeydew}{rgb}{0.94,1.00,0.94}
\definecolor{hotpink}{rgb}{1.00,0.41,0.71}
\definecolor{indianred}{rgb}{0.80,0.36,0.36}
\definecolor{ivory1}{rgb}{1.00,1.00,0.94}
\definecolor{ivory2}{rgb}{0.93,0.93,0.88}
\definecolor{ivory3}{rgb}{0.80,0.80,0.76}
\definecolor{ivory4}{rgb}{0.55,0.55,0.51}
\definecolor{ivory}{rgb}{1.00,1.00,0.94}
\definecolor{khaki1}{rgb}{1.00,0.96,0.56}
\definecolor{khaki2}{rgb}{0.93,0.90,0.52}
\definecolor{khaki3}{rgb}{0.80,0.78,0.45}
\definecolor{khaki4}{rgb}{0.55,0.53,0.31}
\definecolor{khaki}{rgb}{0.94,0.90,0.55}
\definecolor{lavenderblush}{rgb}{1.00,0.94,0.96}
\definecolor{lavender}{rgb}{0.90,0.90,0.98}
\definecolor{lawngreen}{rgb}{0.49,0.99,0.00}
\definecolor{lemonchiffon}{rgb}{1.00,0.98,0.80}
\definecolor{lightblue}{rgb}{0.68,0.85,0.90}
\definecolor{lightcoral}{rgb}{0.94,0.50,0.50}
\definecolor{lightcyan}{rgb}{0.88,1.00,1.00}
\definecolor{lightgoldenrod}{rgb}{0.93,0.87,0.51}
\definecolor{lightgoldenrod}{rgb}{0.98,0.98,0.82}
\definecolor{lightgray}{rgb}{0.83,0.83,0.83}
\definecolor{lightgreen}{rgb}{0.56,0.93,0.56}
\definecolor{lightgrey}{rgb}{0.83,0.83,0.83}
\definecolor{lightpink}{rgb}{1.00,0.71,0.76}
\definecolor{lightsalmon}{rgb}{1.00,0.63,0.48}
\definecolor{lightsea}{rgb}{0.13,0.70,0.67}
\definecolor{lightsky}{rgb}{0.53,0.81,0.98}
\definecolor{lightslate}{rgb}{0.47,0.53,0.60}
\definecolor{lightslate}{rgb}{0.47,0.53,0.60}
\definecolor{lightslate}{rgb}{0.52,0.44,1.00}
\definecolor{lightsteel}{rgb}{0.69,0.77,0.87}
\definecolor{lightyellow}{rgb}{1.00,1.00,0.88}
\definecolor{limegreen}{rgb}{0.20,0.80,0.20}
\definecolor{linen}{rgb}{0.98,0.94,0.90}
\definecolor{magenta1}{rgb}{1.00,0.00,1.00}
\definecolor{magenta2}{rgb}{0.93,0.00,0.93}
\definecolor{magenta3}{rgb}{0.80,0.00,0.80}
\definecolor{magenta4}{rgb}{0.55,0.00,0.55}
\definecolor{magenta}{rgb}{1.00,0.00,1.00}
\definecolor{maroon1}{rgb}{1.00,0.20,0.70}
\definecolor{maroon2}{rgb}{0.93,0.19,0.65}
\definecolor{maroon3}{rgb}{0.80,0.16,0.56}
\definecolor{maroon4}{rgb}{0.55,0.11,0.38}
\definecolor{maroon}{rgb}{0.69,0.19,0.38}
\definecolor{mediumaquamarine}{rgb}{0.40,0.80,0.67}
\definecolor{mediumblue}{rgb}{0.00,0.00,0.80}
\definecolor{mediumorchid}{rgb}{0.73,0.33,0.83}
\definecolor{mediumpurple}{rgb}{0.58,0.44,0.86}
\definecolor{mediumsea}{rgb}{0.24,0.70,0.44}
\definecolor{mediumslate}{rgb}{0.48,0.41,0.93}
\definecolor{mediumspring}{rgb}{0.00,0.98,0.60}
\definecolor{mediumturquoise}{rgb}{0.28,0.82,0.80}
\definecolor{mediumviolet}{rgb}{0.78,0.08,0.52}
\definecolor{midnightblue}{rgb}{0.10,0.10,0.44}
\definecolor{mintcream}{rgb}{0.96,1.00,0.98}
\definecolor{mistyrose}{rgb}{1.00,0.89,0.88}
\definecolor{moccasin}{rgb}{1.00,0.89,0.71}
\definecolor{navajowhite}{rgb}{1.00,0.87,0.68}
\definecolor{navyblue}{rgb}{0.00,0.00,0.50}
\definecolor{navy}{rgb}{0.00,0.00,0.50}
\definecolor{oldlace}{rgb}{0.99,0.96,0.90}
\definecolor{olivedrab}{rgb}{0.42,0.56,0.14}
\definecolor{orange1}{rgb}{1.00,0.65,0.00}
\definecolor{orange2}{rgb}{0.93,0.60,0.00}
\definecolor{orange3}{rgb}{0.80,0.52,0.00}
\definecolor{orange4}{rgb}{0.55,0.35,0.00}
\definecolor{orangered}{rgb}{1.00,0.27,0.00}
\definecolor{orange}{rgb}{1.00,0.65,0.00}
\definecolor{orchid1}{rgb}{1.00,0.51,0.98}
\definecolor{orchid2}{rgb}{0.93,0.48,0.91}
\definecolor{orchid3}{rgb}{0.80,0.41,0.79}
\definecolor{orchid4}{rgb}{0.55,0.28,0.54}
\definecolor{orchid}{rgb}{0.85,0.44,0.84}
\definecolor{palegoldenrod}{rgb}{0.93,0.91,0.67}
\definecolor{palegreen}{rgb}{0.60,0.98,0.60}
\definecolor{paleturquoise}{rgb}{0.69,0.93,0.93}
\definecolor{paleviolet}{rgb}{0.86,0.44,0.58}
\definecolor{papayawhip}{rgb}{1.00,0.94,0.84}
\definecolor{peachpuff}{rgb}{1.00,0.85,0.73}
\definecolor{peru}{rgb}{0.80,0.52,0.25}
\definecolor{pink1}{rgb}{1.00,0.71,0.77}
\definecolor{pink2}{rgb}{0.93,0.66,0.72}
\definecolor{pink3}{rgb}{0.80,0.57,0.62}
\definecolor{pink4}{rgb}{0.55,0.39,0.42}
\definecolor{pink}{rgb}{1.00,0.75,0.80}
\definecolor{plum1}{rgb}{1.00,0.73,1.00}
\definecolor{plum2}{rgb}{0.93,0.68,0.93}
\definecolor{plum3}{rgb}{0.80,0.59,0.80}
\definecolor{plum4}{rgb}{0.55,0.40,0.55}
\definecolor{plum}{rgb}{0.87,0.63,0.87}
\definecolor{powderblue}{rgb}{0.69,0.88,0.90}
\definecolor{purple1}{rgb}{0.61,0.19,1.00}
\definecolor{purple2}{rgb}{0.57,0.17,0.93}
\definecolor{purple3}{rgb}{0.49,0.15,0.80}
\definecolor{purple4}{rgb}{0.33,0.10,0.55}
\definecolor{purple}{rgb}{0.63,0.13,0.94}
\definecolor{red1}{rgb}{1.00,0.00,0.00}
\definecolor{red2}{rgb}{0.93,0.00,0.00}
\definecolor{red3}{rgb}{0.80,0.00,0.00}
\definecolor{red4}{rgb}{0.55,0.00,0.00}
\definecolor{red}{rgb}{1.00,0.00,0.00}
\definecolor{rosybrown}{rgb}{0.74,0.56,0.56}
\definecolor{royalblue}{rgb}{0.25,0.41,0.88}
\definecolor{saddlebrown}{rgb}{0.55,0.27,0.07}
\definecolor{salmon1}{rgb}{1.00,0.55,0.41}
\definecolor{salmon2}{rgb}{0.93,0.51,0.38}
\definecolor{salmon3}{rgb}{0.80,0.44,0.33}
\definecolor{salmon4}{rgb}{0.55,0.30,0.22}
\definecolor{salmon}{rgb}{0.98,0.50,0.45}
\definecolor{sandybrown}{rgb}{0.96,0.64,0.38}
\definecolor{seagreen}{rgb}{0.18,0.55,0.34}
\definecolor{seashell1}{rgb}{1.00,0.96,0.93}
\definecolor{seashell2}{rgb}{0.93,0.90,0.87}
\definecolor{seashell3}{rgb}{0.80,0.77,0.75}
\definecolor{seashell4}{rgb}{0.55,0.53,0.51}
\definecolor{seashell}{rgb}{1.00,0.96,0.93}
\definecolor{sienna1}{rgb}{1.00,0.51,0.28}
\definecolor{sienna2}{rgb}{0.93,0.47,0.26}
\definecolor{sienna3}{rgb}{0.80,0.41,0.22}
\definecolor{sienna4}{rgb}{0.55,0.28,0.15}
\definecolor{sienna}{rgb}{0.63,0.32,0.18}
\definecolor{skyblue}{rgb}{0.53,0.81,0.92}
\definecolor{slateblue}{rgb}{0.42,0.35,0.80}
\definecolor{slategray}{rgb}{0.44,0.50,0.56}
\definecolor{slategrey}{rgb}{0.44,0.50,0.56}
\definecolor{snow1}{rgb}{1.00,0.98,0.98}
\definecolor{snow2}{rgb}{0.93,0.91,0.91}
\definecolor{snow3}{rgb}{0.80,0.79,0.79}
\definecolor{snow4}{rgb}{0.55,0.54,0.54}
\definecolor{snow}{rgb}{1.00,0.98,0.98}
\definecolor{springgreen}{rgb}{0.00,1.00,0.50}
\definecolor{steelblue}{rgb}{0.27,0.51,0.71}
\definecolor{tan1}{rgb}{1.00,0.65,0.31}
\definecolor{tan2}{rgb}{0.93,0.60,0.29}
\definecolor{tan3}{rgb}{0.80,0.52,0.25}
\definecolor{tan4}{rgb}{0.55,0.35,0.17}
\definecolor{tan}{rgb}{0.82,0.71,0.55}
\definecolor{thistle1}{rgb}{1.00,0.88,1.00}
\definecolor{thistle2}{rgb}{0.93,0.82,0.93}
\definecolor{thistle3}{rgb}{0.80,0.71,0.80}
\definecolor{thistle4}{rgb}{0.55,0.48,0.55}
\definecolor{thistle}{rgb}{0.85,0.75,0.85}
\definecolor{tomato1}{rgb}{1.00,0.39,0.28}
\definecolor{tomato2}{rgb}{0.93,0.36,0.26}
\definecolor{tomato3}{rgb}{0.80,0.31,0.22}
\definecolor{tomato4}{rgb}{0.55,0.21,0.15}
\definecolor{tomato}{rgb}{1.00,0.39,0.28}
\definecolor{turquoise1}{rgb}{0.00,0.96,1.00}
\definecolor{turquoise2}{rgb}{0.00,0.90,0.93}
\definecolor{turquoise3}{rgb}{0.00,0.77,0.80}
\definecolor{turquoise4}{rgb}{0.00,0.53,0.55}
\definecolor{turquoise}{rgb}{0.25,0.88,0.82}
\definecolor{violetred}{rgb}{0.82,0.13,0.56}
\definecolor{violet}{rgb}{0.93,0.51,0.93}
\definecolor{wheat1}{rgb}{1.00,0.91,0.73}
\definecolor{wheat2}{rgb}{0.93,0.85,0.68}
\definecolor{wheat3}{rgb}{0.80,0.73,0.59}
\definecolor{wheat4}{rgb}{0.55,0.49,0.40}
\definecolor{wheat}{rgb}{0.96,0.87,0.70}
\definecolor{whitesmoke}{rgb}{0.96,0.96,0.96}
\definecolor{white}{rgb}{1.00,1.00,1.00}
\definecolor{yellow1}{rgb}{1.00,1.00,0.00}
\definecolor{yellow2}{rgb}{0.93,0.93,0.00}
\definecolor{yellow3}{rgb}{0.80,0.80,0.00}
\definecolor{yellow4}{rgb}{0.55,0.55,0.00}
\definecolor{yellowgreen}{rgb}{0.60,0.80,0.20}
\definecolor{yellow}{rgb}{1.00,1.00,0.00}

\newcommand{\red}[1]{{\color{red}#1}}

\def\m#1{\mathcal#1}

\newcommand{\be}{\begin{equation}}
\newcommand{\ee}{\end{equation}}

\def\bea{\begin{eqnarray}}
\def\eea{\end{eqnarray}}
\def\bean{\begin{eqnarray*}}
\def\eean{\end{eqnarray*}}

\def\cl{\centerline}

\def\head#1{$$gin{center}
\newcommand{\group}{$\mathcal{ PSL}_2(7)$ }
\newcommand{\more}{\vskip .2cm\hskip 4cm{\bf TO BE CONTINUED}\vskip .2cm}
\newcommand{\ligne}{\vskip .5cm\noindent}
\def\ov{\overline}
\shadowbox{
\begin{minipage}{3.0in}
\begin{center}
\bf \red{#1}
\end{center}
\end{minipage}}\end{center}\vspace{0.5\semcm} }
\begin{document}




\title{\hfill ~\\[-30mm]
       \hfill\mbox{\small }\\[30mm]
       \textbf{ Neutrinos and Particle Physics Models}
       } 
\date{}
\author{\\ Pierre Ramond\footnote{E-mail: {\tt pierre.ramond@gmail.com}} \footnote{\tt Talk presented at ``History of the Neutrino" Paris,  September 2018}\\ \\
  \emph{\small{}Institute for Fundamental Theory, Department of Physics,}\\
  \emph{\small University of Florida, Gainesville, FL 32611, USA}
  }


\maketitle

\begin{abstract}
\noindent   
As in Greek mythology, the neutrino was born in the mind of Wolfgang Pauli to salvage a fundamental principle. Its existence met with universal skepticism by a scientific community used to infer particles from experiment. Its detection in 1956 brought particle physics acceptance; its chirality explained maximal parity violation in $\beta$ decay; its (apparent) masslessness led theorists to imagine new symmetries. Neutrinos are pioneers of mutli-messenger astronomy, from the Sun, from SNA1987, and now through IceCube's blazar. The discovery of neutrino masses opened a new era in particle physics aswell as  unexplored windows on the universe. -Tiny neutrino masses suggest new physics at very short distances through the Seesaw.
- Neutrinos and quarks, unified by gauge structure, display different mass and mixing patterns: small quark mixing angles and two large neutrino mixing angles. This difference in mass and mixings in the midst of gauge unification may be an important clue towards Yukawa unification.
- Neutrino mixings provide a new source of CP-violation, and may solve the riddle of matter-antimatter asymmetry. We present a historical journey of  these ``enfants terribles" of particle physics and their importance in understanding our universe.

\end{abstract} 

\thispagestyle{empty}
\vfill
\newpage
\setcounter{page}{1}

\section{Preamble}
When asked my occupation in life, I often answer that I study neutrinos.  My attempts at elaboration motivated an artist  acquaintance  to produce these visual portaits of neutrinos,

\vskip .5cm
\hskip .5in
\includegraphics[scale=.2]{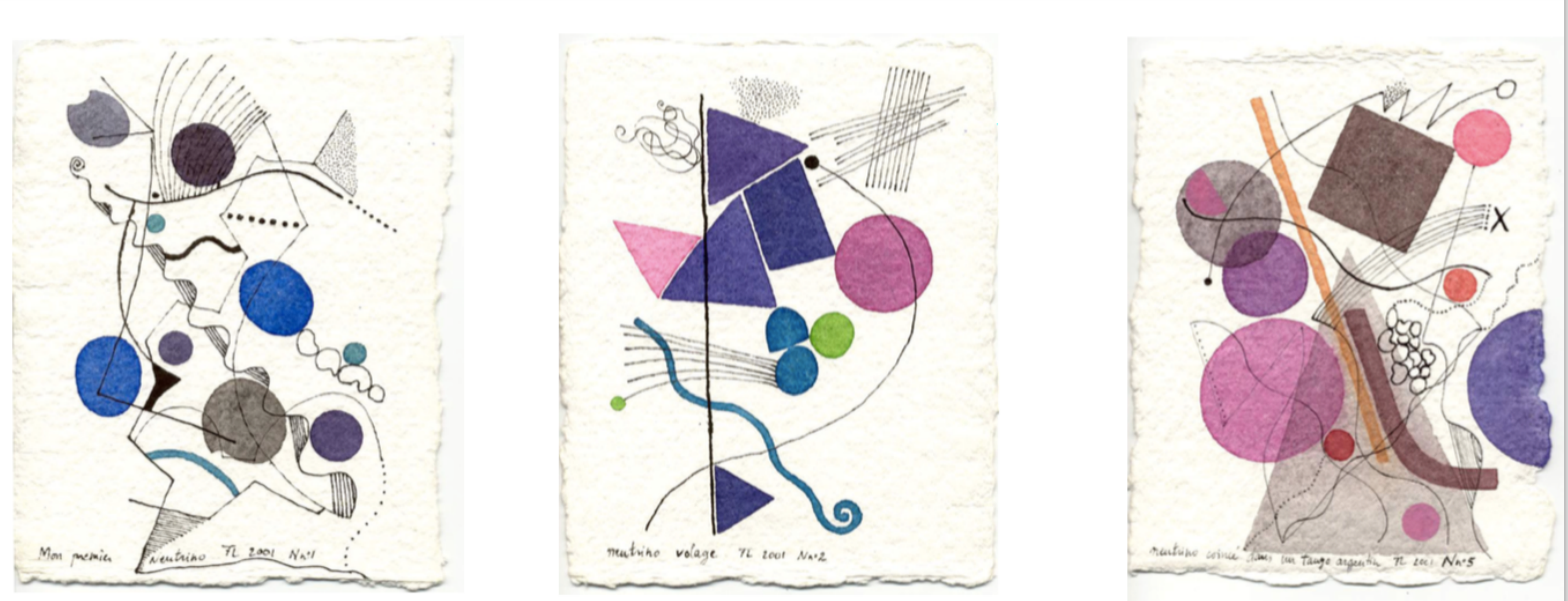}
\vskip .5cm
\noindent  On my home office wall  they remind me of the evocative powers of neutrinos on our imagination. 
\vskip .5cm
\noindent
This talk consists of four parts:
\begin{itemize}
\item{\large Editorial}

\item{\large Early History}

\item{\large Neutrino Masses}

\item{\large Neutrinos \& Yukawa Unification}
\end{itemize}
 
\section{Editorial}      
 The idea of a neutrino was revealed to Wolfgang Pauli, not through direct experimental evidence but as a ``desperate" attempt to rescue what he believed to be a fundamental principle: the conservation of energy. He was right, of course, but Pauli's neutron (neutrino) was difficult if not impossible to detect, and for a while he lamented on his fate, having invented a particle impossible to detect\footnote{not unlike the axion?}. In his days, inventing a new particle seemed like an admission of failure, to be contrasted with the present sociology where a mere glitch in the data generates a whole Kaluza-Kein tower of particles!
  
 For experimentalists (and most theorists) his hypothesis was not taken seriously at first, even though his proposal added a spin one half particle in the nucleus, thereby explaining in addition the intensity of Raman lines from the Nitrogen nucleus. 
    
This disrespect of the neutrino concept was surely misplaced as neutrinos are the misfits of the particle world; they never fit current dogma. Retrospectively,
\vskip .5cm\noindent 
- Neutrinos are left-handed in an ambidextruous world, generating parity violation in $\beta$ decay.

\vskip .5cm\noindent
- Neutrinos appeared to be massless, motivating theorists to seek a general principle for their lack of mass; witness Volkov and Akulov's non-linear representation of supersymmetry with the neutrino as Nambu-Goldstone fermion, and Fayet's proposal of a supersymmetric Standard Model.
\vskip .5cm\noindent
- Neutrinos may be Majorana particles, leading to  leptogenesis and possibly explaining matter-antimatter asymmetry.
\vskip .5cm\noindent
- Absurdly light neutrinos require a new scale of physics?
\vskip .5cm\noindent
- Neutrinos as keys to Yukawa Unification: they display the  same gauge structure as quarks, yet their Yukawa patterns are strikingly  different. This outstanding problem begs explanation.
 \vskip .5cm\noindent
-  Neutrinos are messengers from the Universe, from the center of the Sun, from Supernovae, and recently detected by IceCube from a  four billion years old blazar! 
 \vskip .5cm\noindent
Except for dark matter, Neutrino masses and mixings provide the only ``Physics Beyond the Standard Model". Today a small proportion of particle physicists work on neutrino, even though over the years a number of  neutrino prospectors found their study very rewarding: 
 
 \vskip .5cm
\hskip -0.25in
\includegraphics[scale=.2]{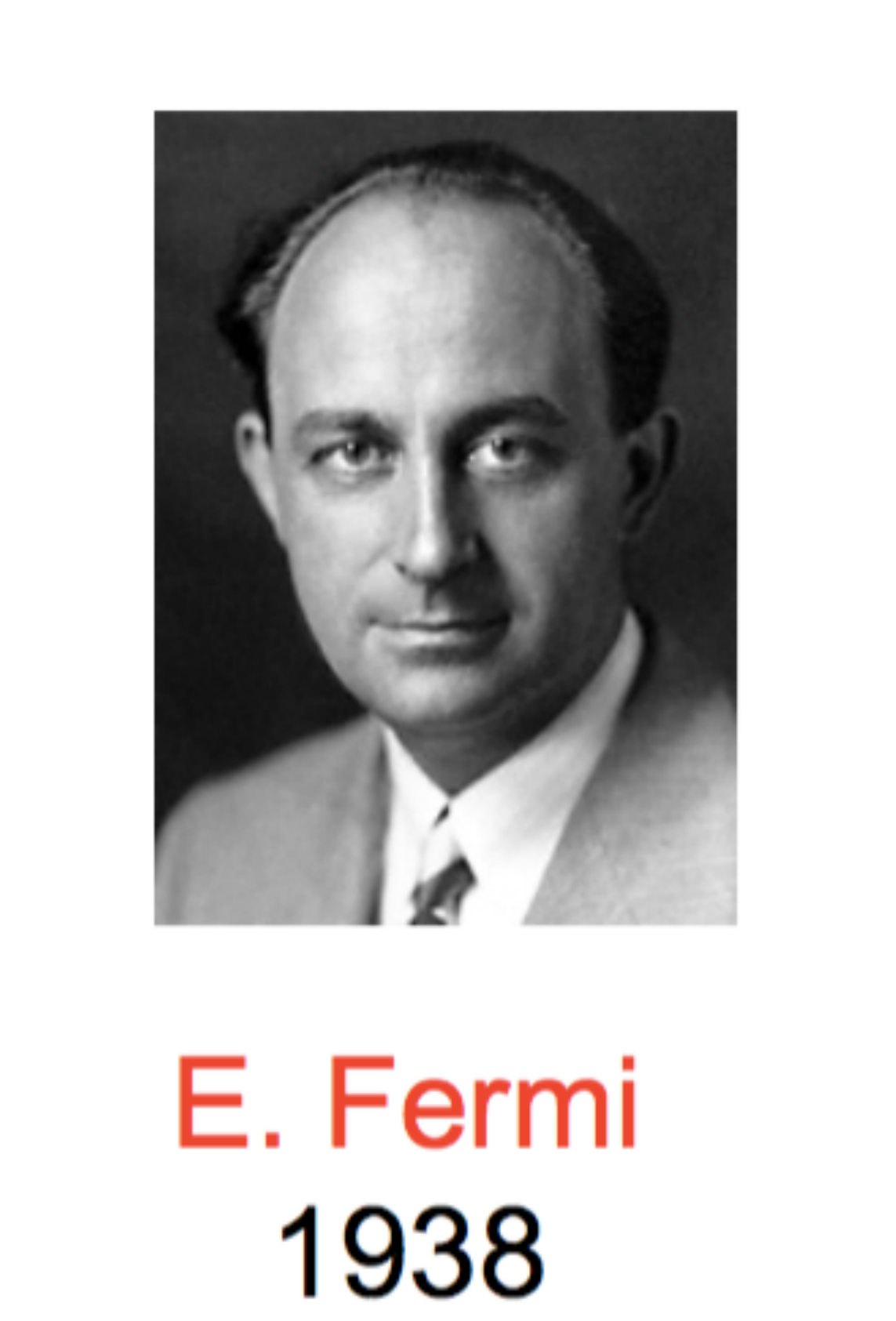}
\hskip 0.2in
\includegraphics[scale=.2]{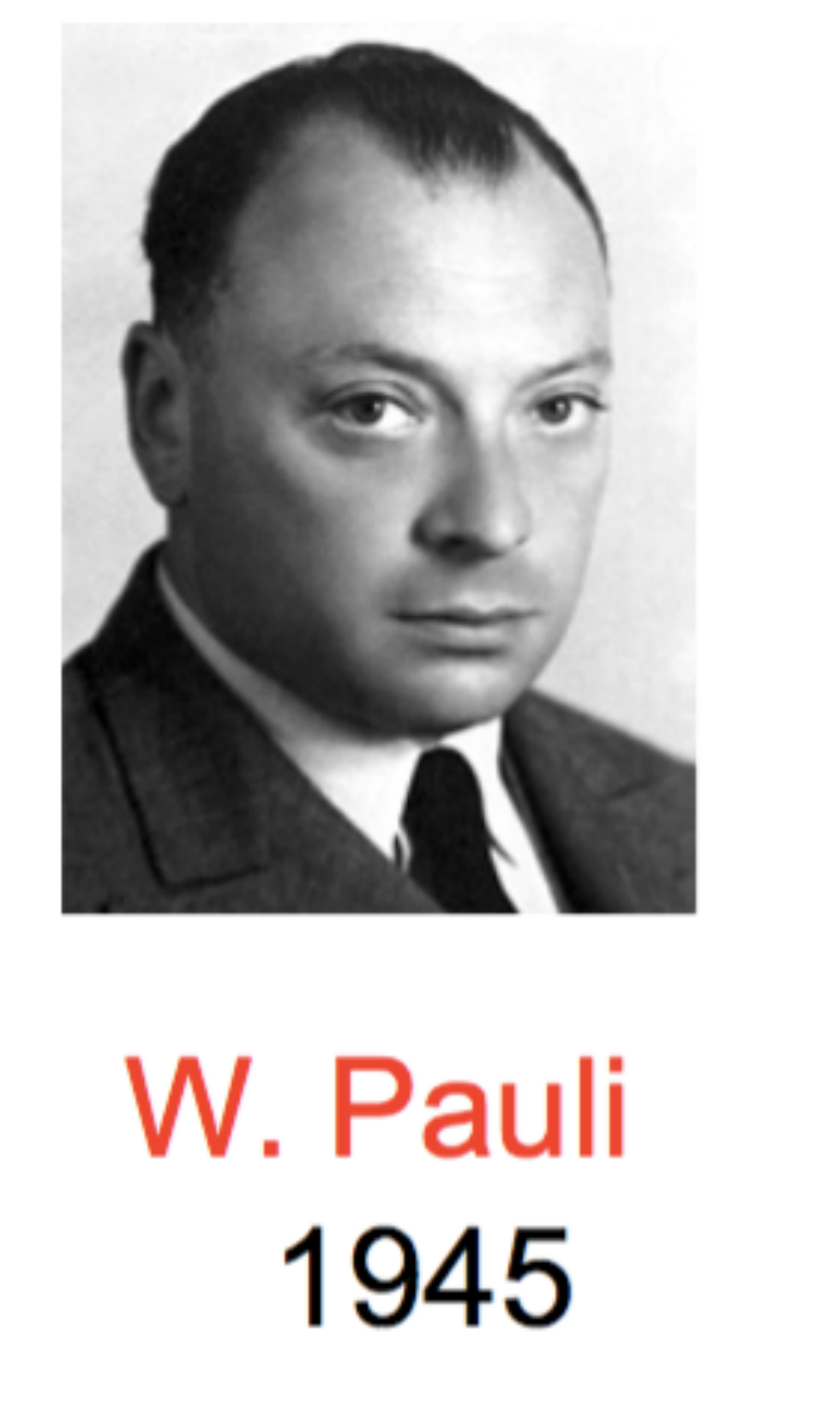}
\hskip 0.2in
\includegraphics[scale=.2]{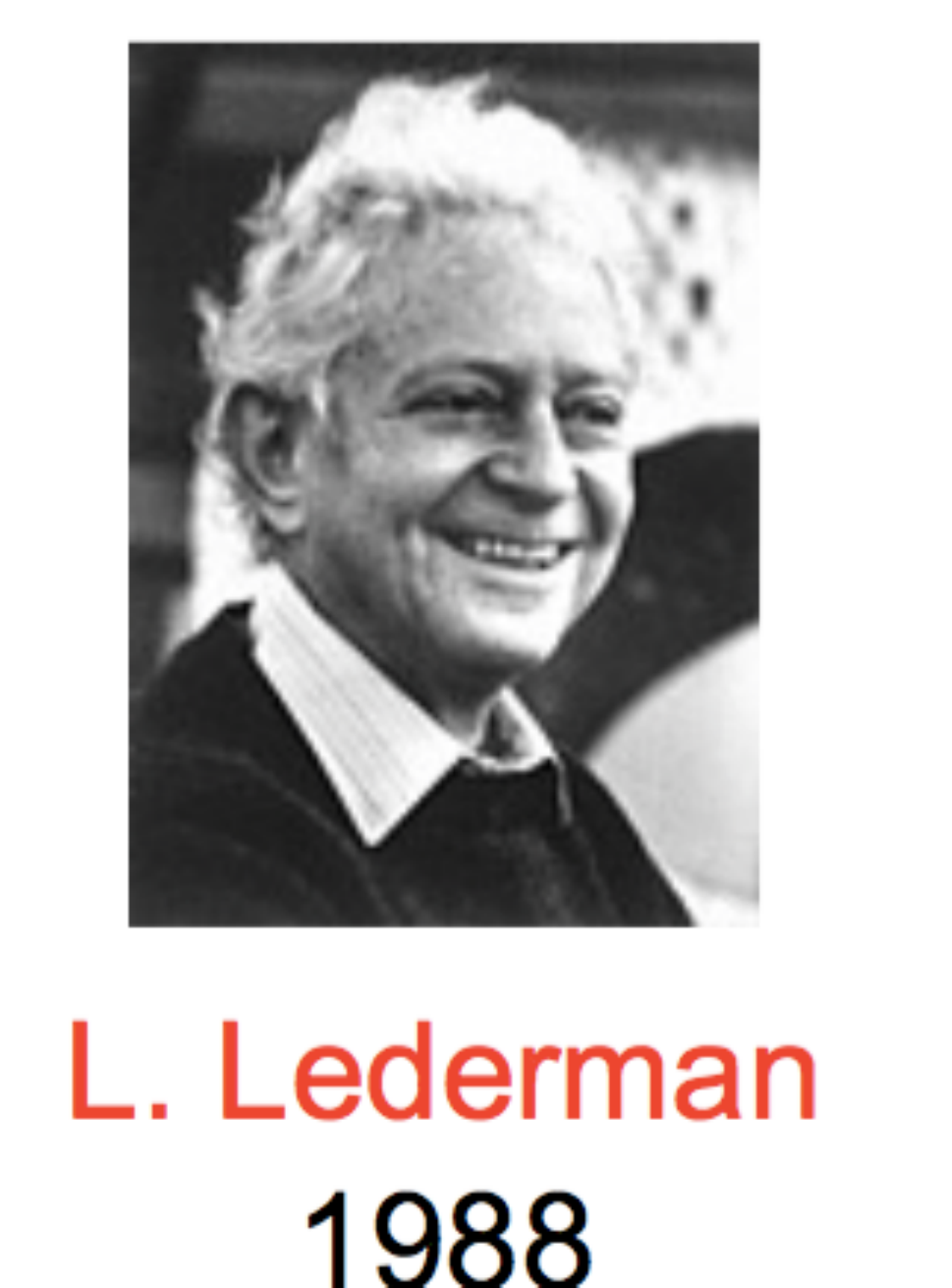}
\hskip 0.2in
\includegraphics[scale=.2]{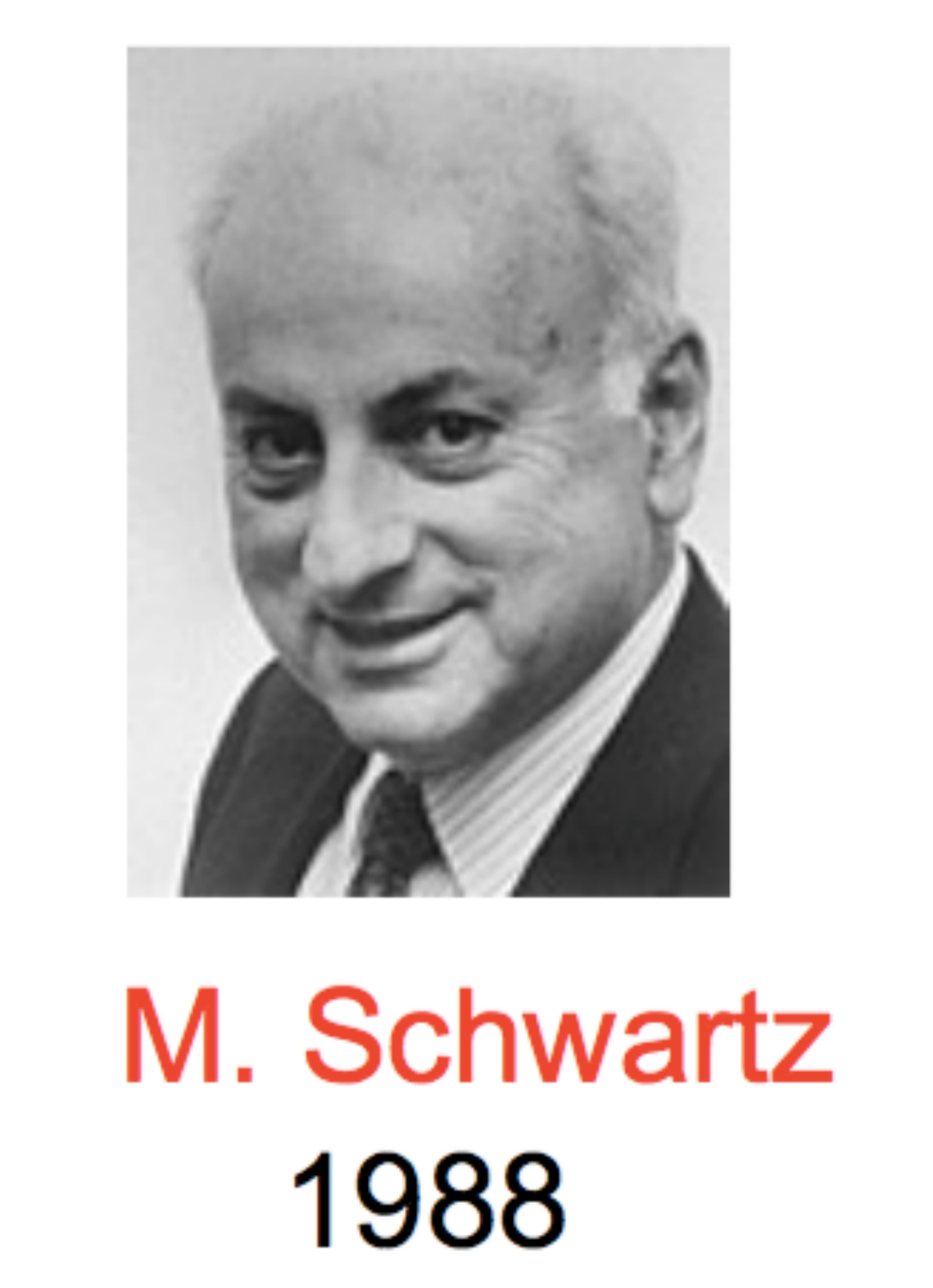}
\hskip 0.2in
\includegraphics[scale=.2]{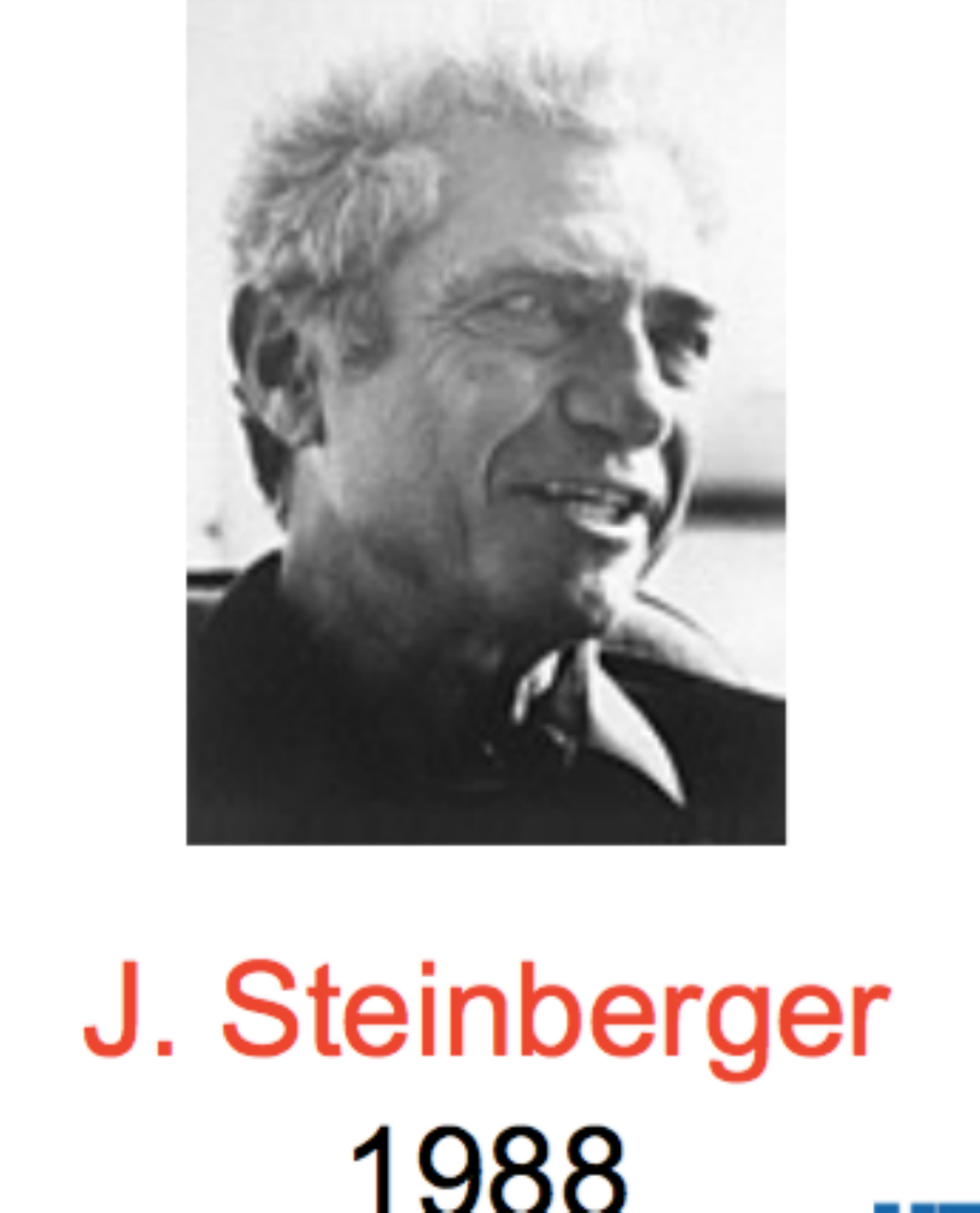}
\vskip .5cm
\hskip -0.25in
\includegraphics[scale=.2]{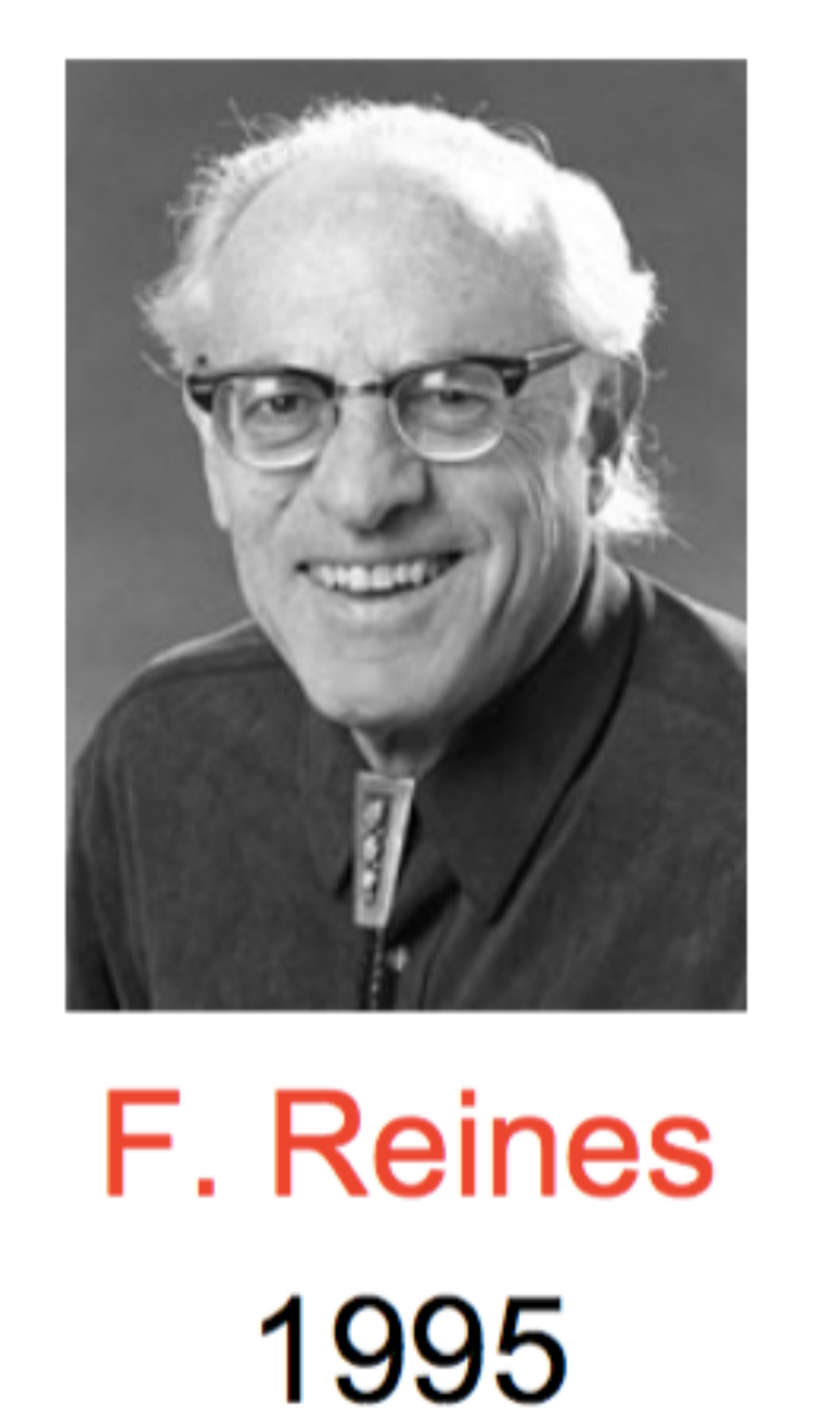}
\hskip 0.2in
\includegraphics[scale=.2]{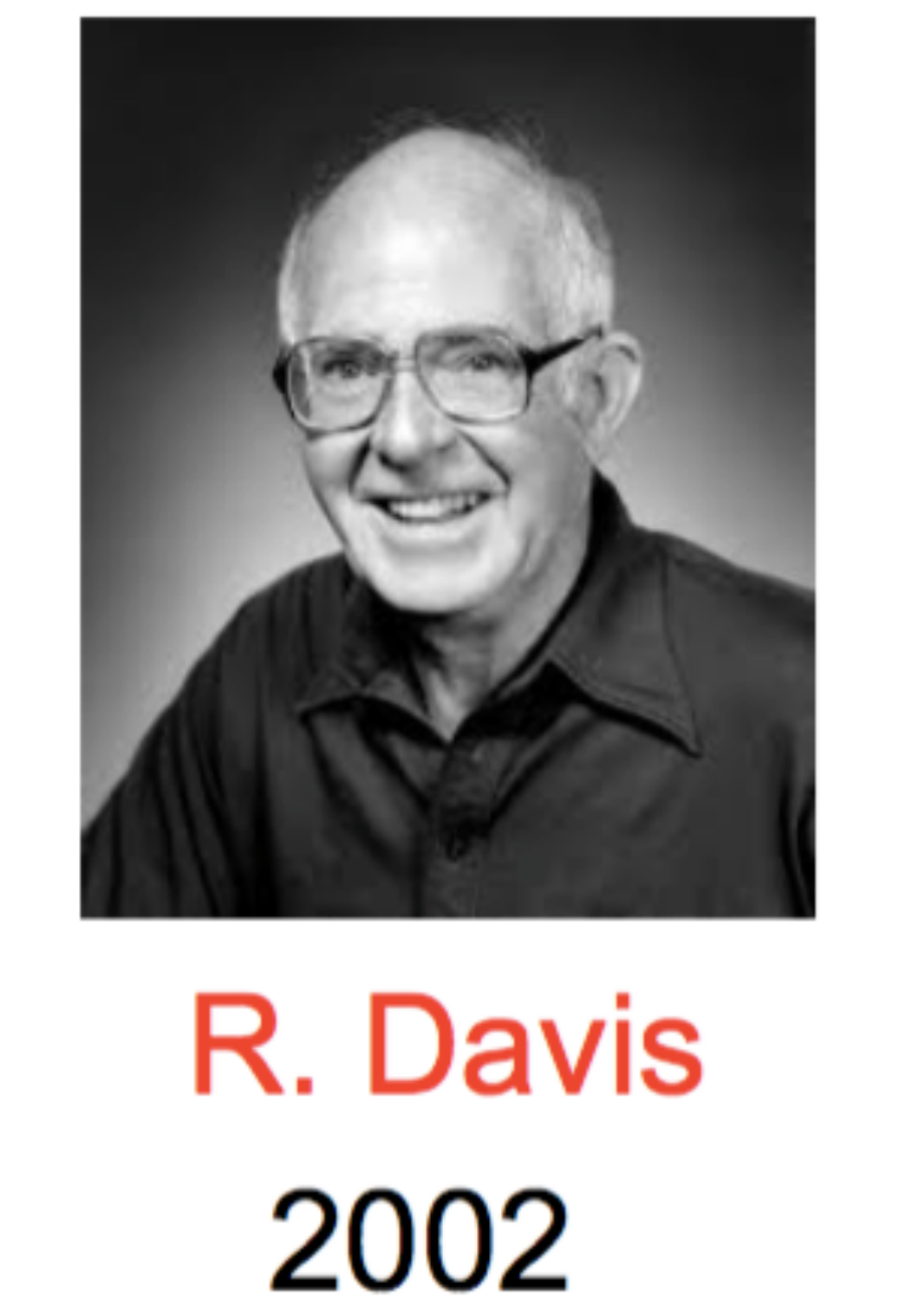}
\hskip 0.2in
\includegraphics[scale=.2]{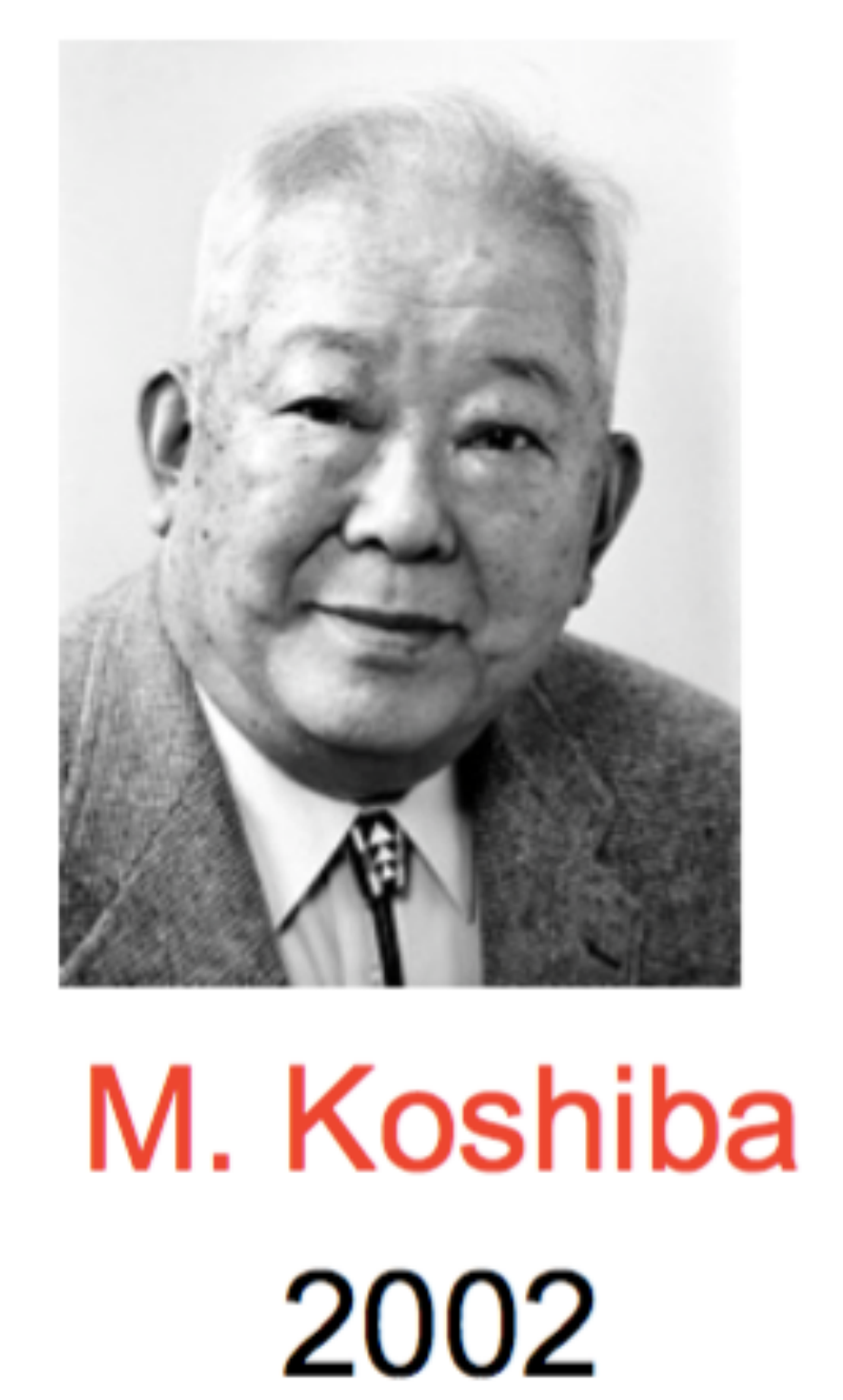}
\hskip 0.2in
\includegraphics[scale=.2]{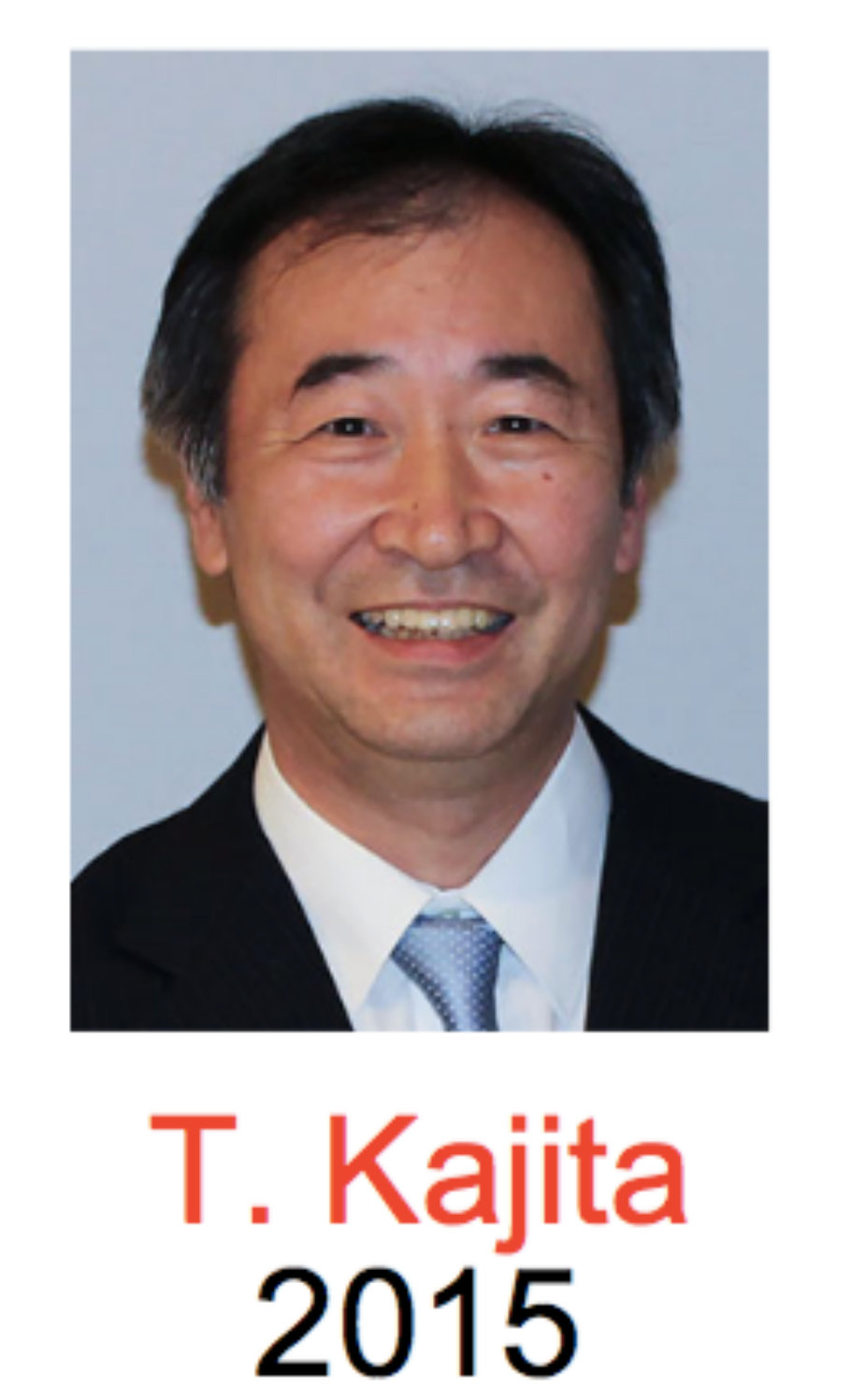}
\hskip 0.2in
\includegraphics[scale=.2]{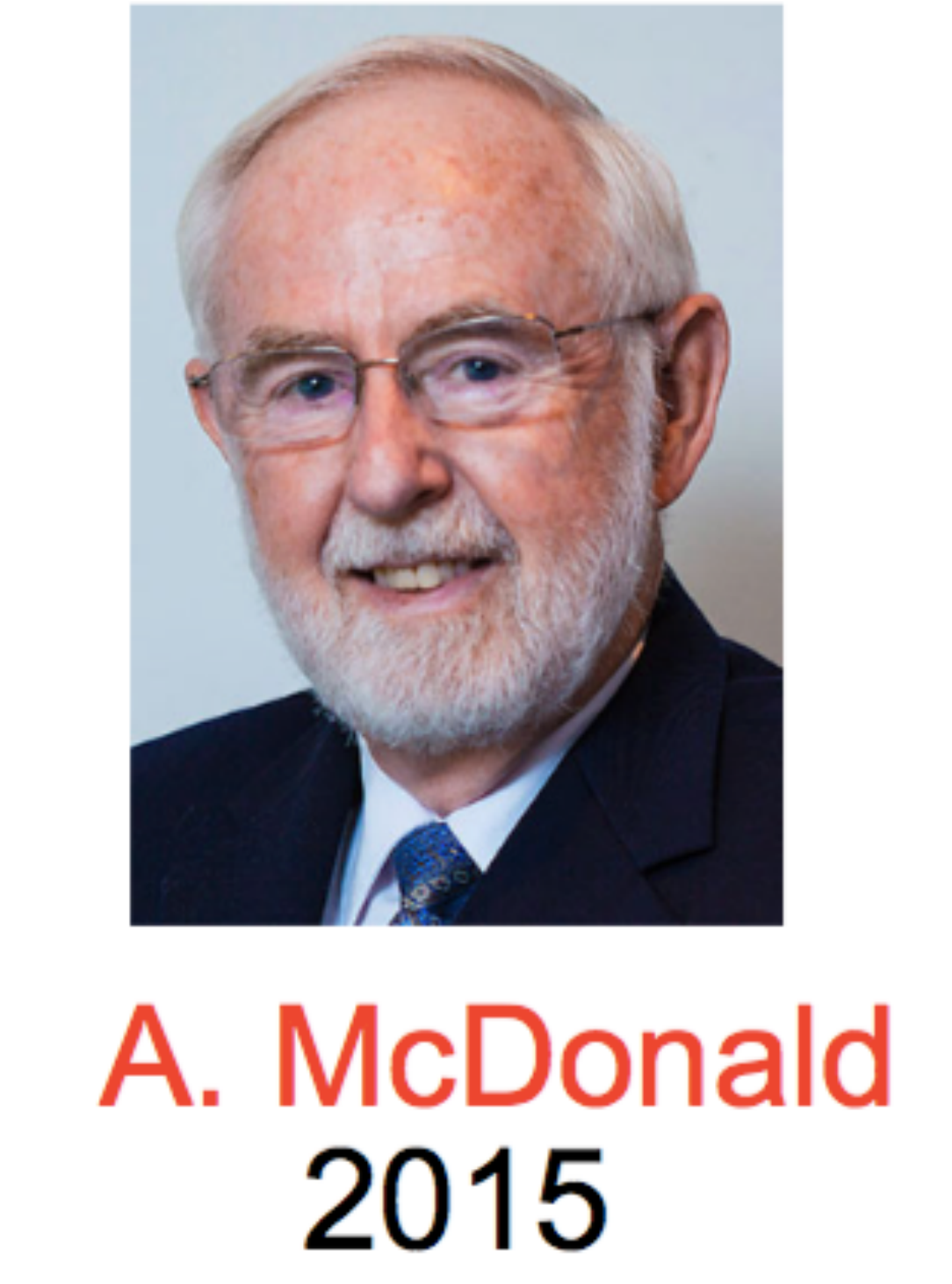}
\vskip .5cm
\noindent 
Not to mention those notables who belong to  the Neutrino Hall of Fame 
 \vskip 1cm
 \hskip 0.2in
\includegraphics[scale=.25]{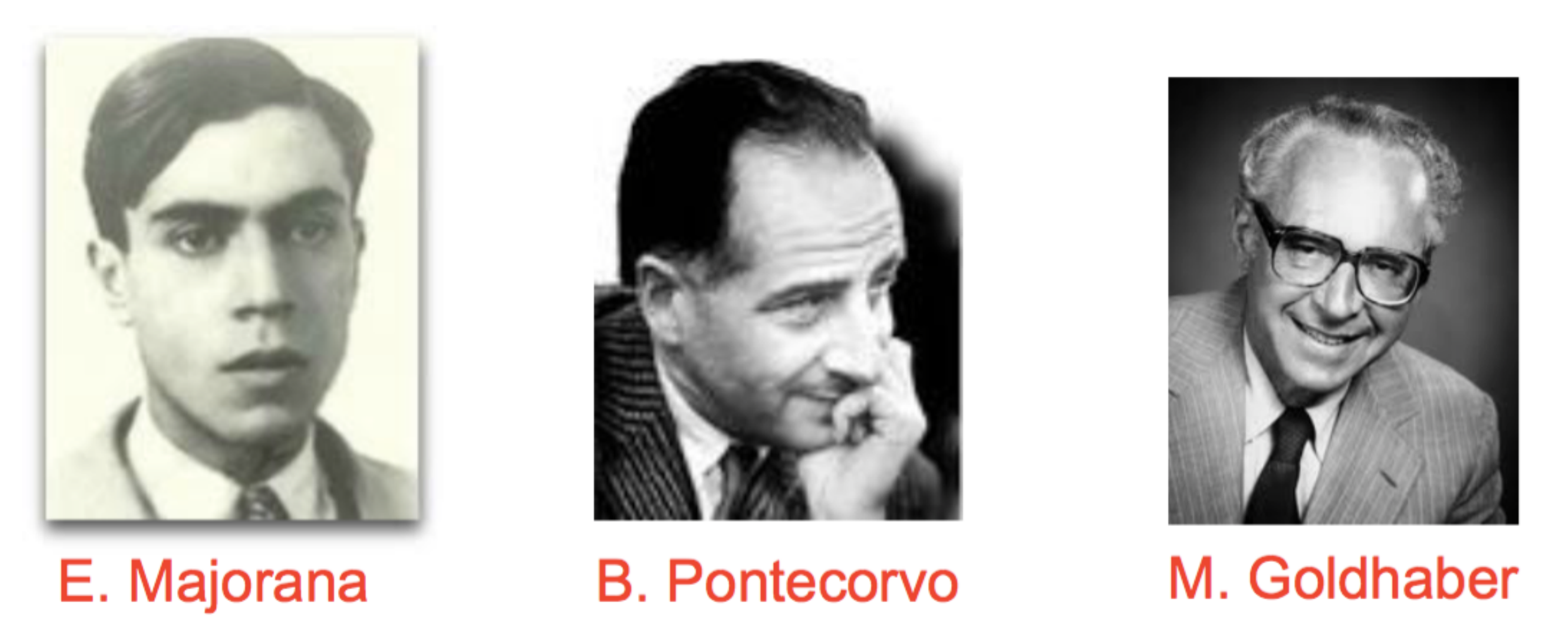}
\vskip .5cm 
\hskip 0.2in
\includegraphics[scale=.25]{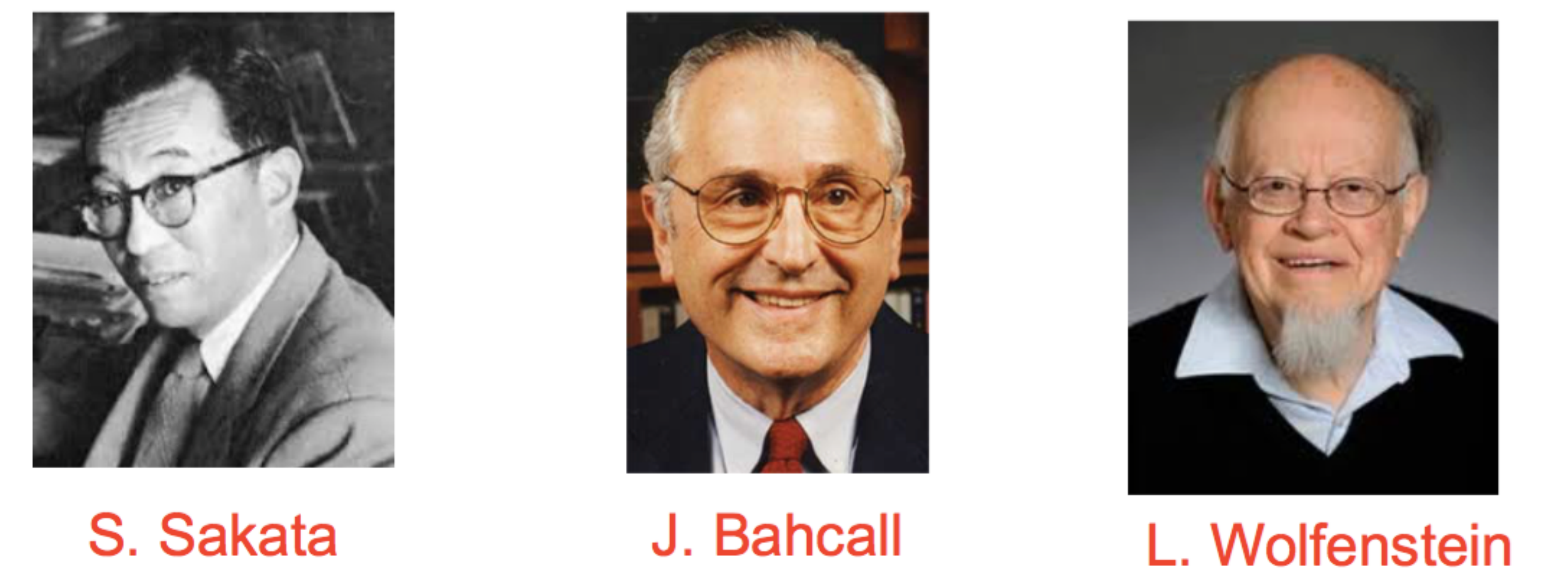}
 
 \vskip .5cm\noindent
Their  past achievements suggest that it may not be a bad idea to study everything possible about neutrinos\footnote{In the absence of direct evidence, theorists  should put wax in their ears  and chain themselves to the mast to resist the lure of light sterile neutrinos, while of course urging experimentalists to look for them}.
  
 \vskip .5cm\noindent
 Enough editorializing, and let us look at the neutrino's  early history

 \newpage
\section{Early History}
It is customary to begin with Pauli's famous letter to Lise Meitner and friends,
\vskip .5cm

\includegraphics[scale=.3]{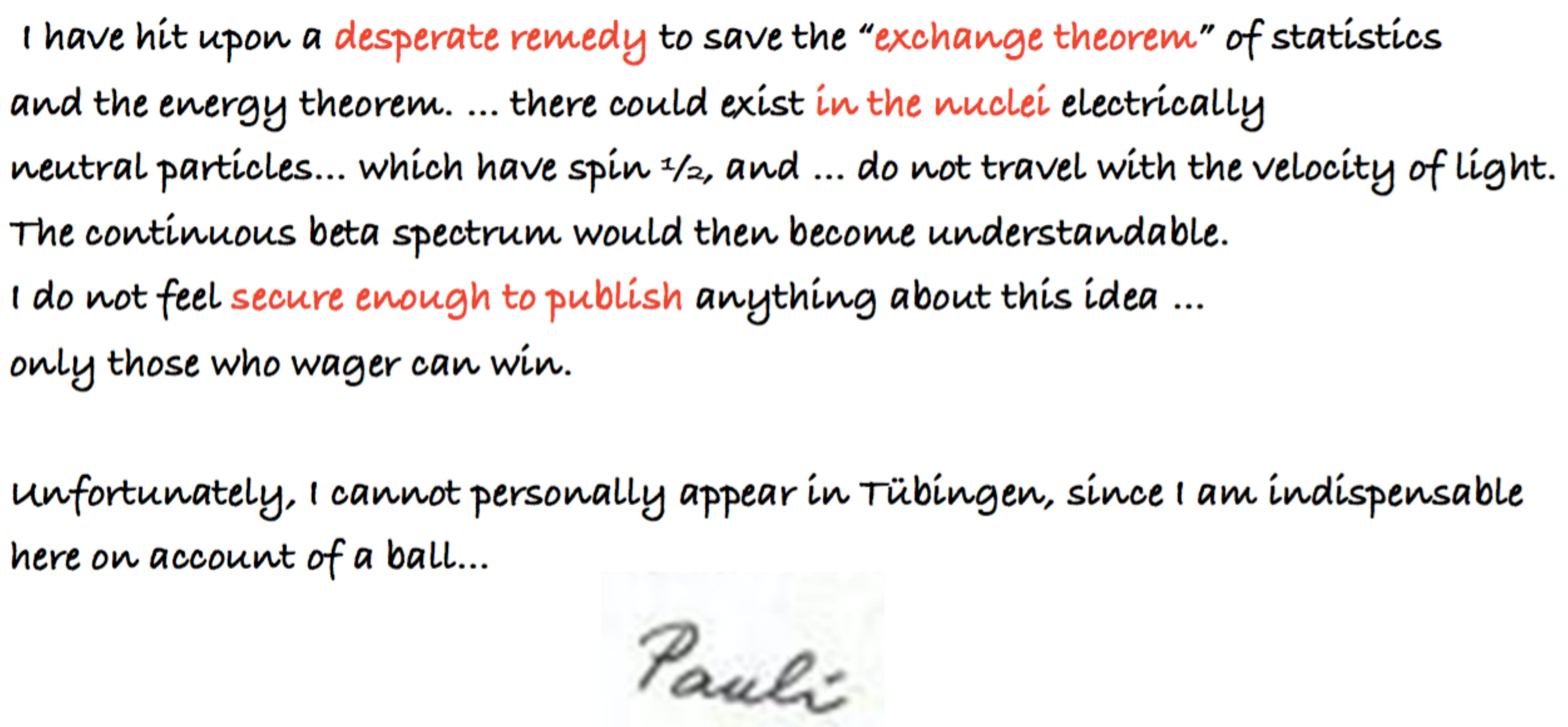}
 \vskip .5cm\noindent
which is noteworthy in many different ways.  Pauli postulates the existence of a neutral particle in the nucleus. Its existence would then solve two experimental facts. Raman scattering of the  Nitrogen nucleus  implies it is a boson.  In Pauli's world, the Nitrogen nucleus is made up of protons and electrons and  to account for its atomic weight and chemistry it must  contain $7+7$ protons and $7$ electrons, thus  making it a fermion. This is the ``exchange theorem" part as a new spin one-half fermion in the nucleus solves that problem. It is only later in the letter that he mentions the continuous spectrum of the $\beta$ electron, and in order to account for his particle to be in the nucleus, he endows it with a magnetic moment, and therefore a mass!
\vskip .5cm\noindent

Chadwick's discovery of neutron two years later solves the Nitrogen problem, and does not require Pauli's light neutron to be inside the nucleus.  However it is still needed, although in a new world rocked by quantum mechanics, even the great Bohr entertained the idea that  nuclear processes might violate energy conservation. 
\vskip .5cm\noindent

The sociological context of the letter is revealing.  Pauli is clearly nervous at the idea of introducing a new particle! So much so that he does not publish the idea. Six months later, at the  APS June 1931 meeting in Pasadena, Pauli gave a talk where he is said to have discussed his new particle and  believed it lived in the nucleus. I have not been able to find a copy of his talk.

\vskip .5cm\noindent
One might wonder if the Neutron had been discovered earlier (as it could have been) would Pauli have suggested a new light neutral particle? Did the founding fathers think that they should solve every puzzle without introducing new degrees of feedom? Contrast with today's practice where any experimental anomaly is interpreted by new particles, even towers thereof. ``O Tempore O Mores".
\vskip .5cm\noindent 
Another aspect of the letter is that he foregoes a physics meeting to go on a date! Pauli was in the midst of a divorce from actress Kate Depner who left him for a chemist! Within a year Pauli was under analysis with Carl Jung.
\vskip .5cm\noindent
It was of course E. Fermi who  in $1933$ and $1934$ papers identifies Pauli's particle as being created by the decay process. Being Italian he named it neutrino, the little neutron.   
 \vskip 1cm
 
 \noindent A revealing  testimony of the place the neutrino idea occupied in particle physics is  Hans Bethe and Robert Bacher's $1936$ Review of Modern Physics 
 \vskip .5cm
 \hskip 1cm \includegraphics[scale=.3]{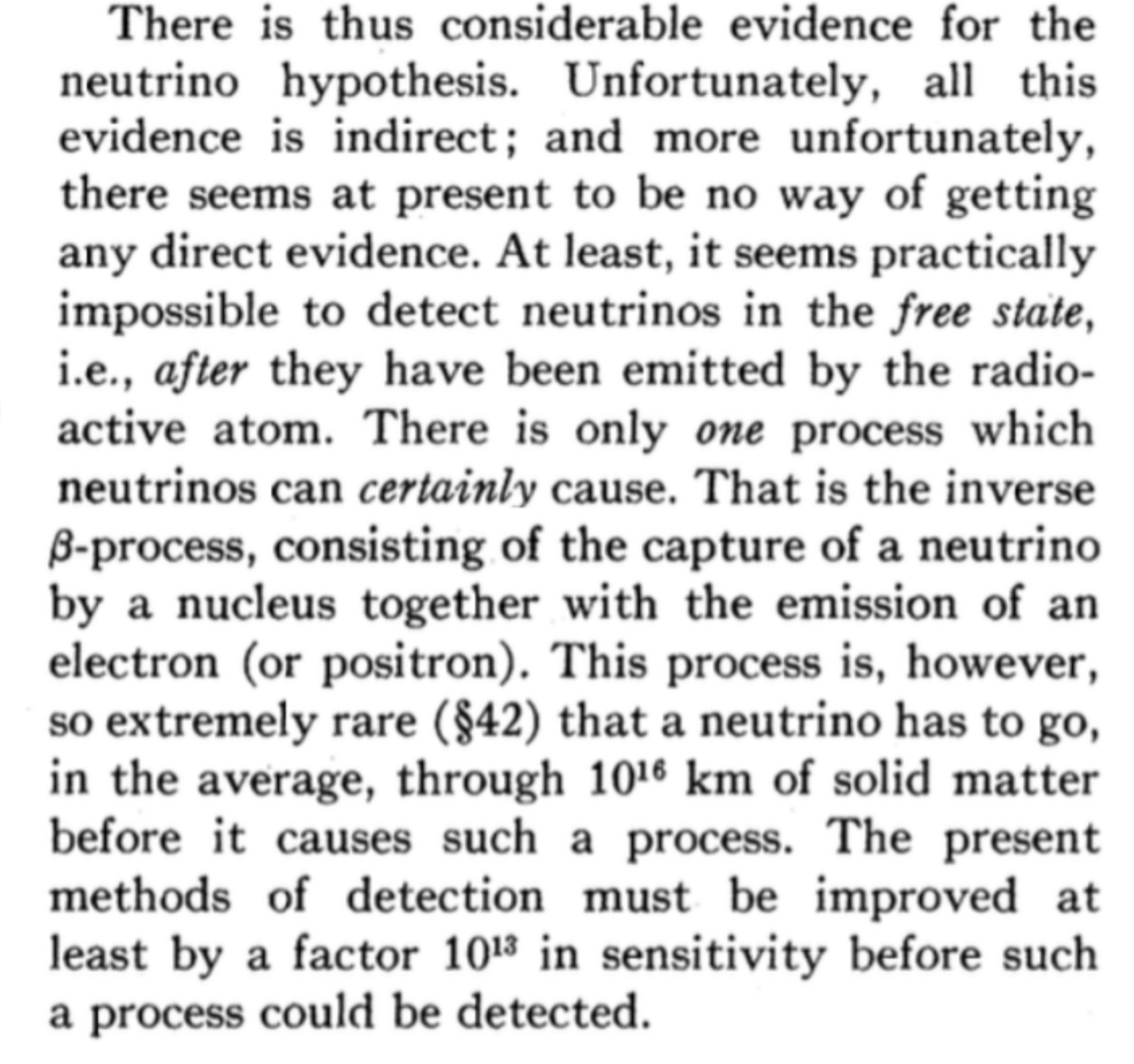}
\vskip .5cm\noindent
Interesting as it may be, the neutrino idea offers no proof of its existence. Still they identify the process by which the (anti)neutrino was detected twenty years later: inverse $\beta$ decay. Its detection required an improvement of $10^{13}$ in sensitivity, making it all but insurmountable!
\vskip .5cm\noindent
Bethe and Bacher still denote the neutrino by $n'$ to distinguish it from the neutron $n$.  L. H. Rumbaugh, R. B. Roberts and L. R. Hafstad seem to be the first to  use the greek letter $\nu$ in $1937$ (E. M. Lyman a year later). I am not aware of any earlier attribution. It is universally used from then on. 
\vskip .5cm\noindent
Ten years later, the $1948$ Reviews of Modern Physics article by H. R. Crane summarizes the community's attitude on the neutrino, as a useful idea but still not universally accepted:

\vskip .5cm
 \hskip 1.5cm \includegraphics[scale=.3]{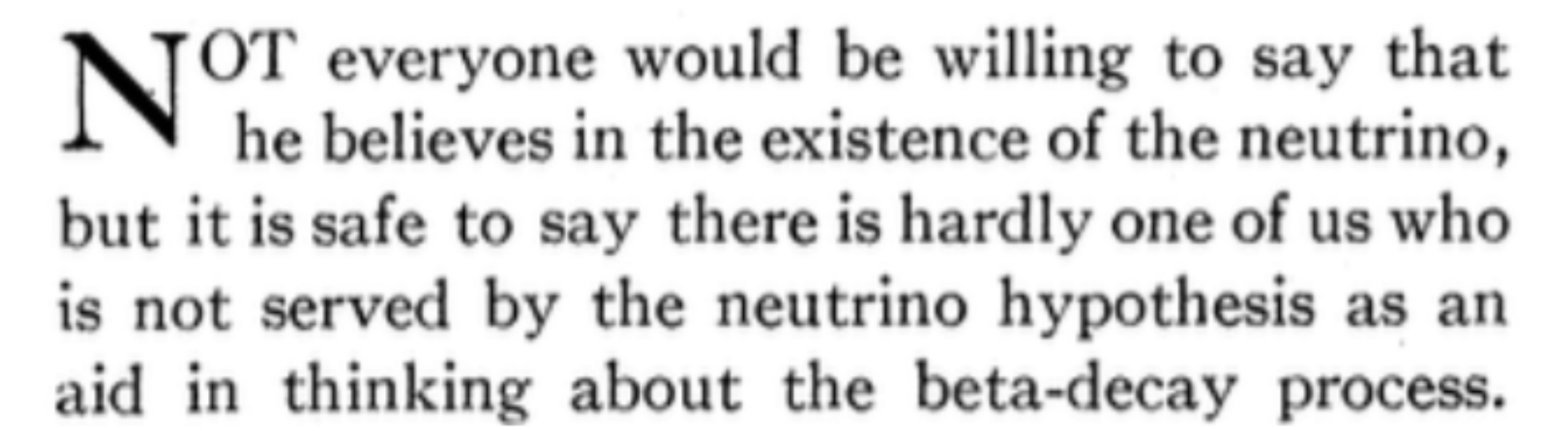}
\vskip 1cm\noindent 
This attitude is about to change when Clyde Cowan and Frederick Reines  use inverse $\beta$ decay to finally detect antineutrinos coming from the Savannah River reactor at the Georgia-South Carolina border. The neutrino is the only elementary particle discovered south of the Mason-Dixon line. At first  their discovery met with skepticism, as the titles of their papers suggest:  $1953$   ``Detection of the Free Neutrino"  announce the experiment,  the $1954$ talk ``Status of an Experiment to detect the free neutrino" at the January APS Meeting, and finally their $1956$ article ``Detection of a Free Neutrino: a Confirmation", published in Nature. Earlier, Cowan and Reines had sent Pauli news of their discovery who responded thus,
\vskip .5cm
 \hskip 1.5cm \includegraphics[scale=.25]{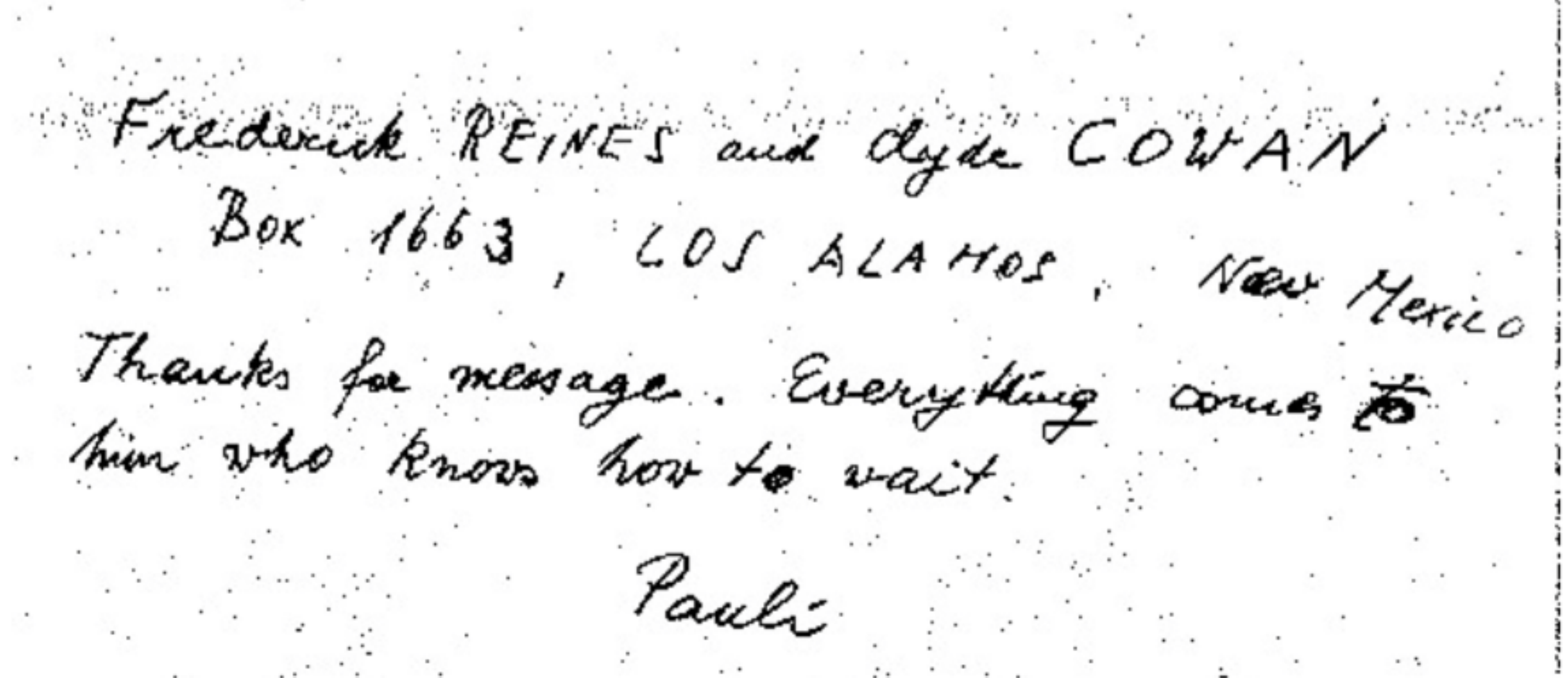}
\vskip .5cm\noindent
A comment very much applicable to the present state of particle physics! 
\vskip 1cm\noindent
In $1937$ E. Majorana (Il Nuovo Cimento 14,171(1937)) noticed that as a neutral particle the neutrino could, without violating Lorentz invariance, be its own antiparticle, in constrast with electrons and positrons easily distinguishable by their electrical charge. 

\vskip .5cm
\hskip 1cm \includegraphics[scale=.25]{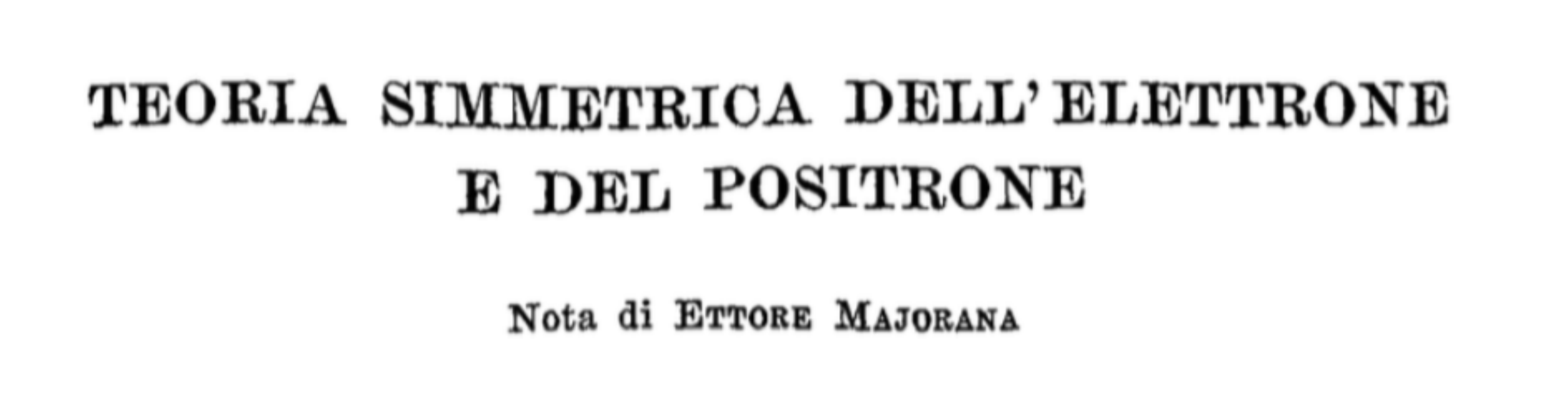}
\vskip .5cm\noindent This brilliant theoretical remark will assume more importance in later years. Neutrinos and antineutrinos can be distinguished by their lepton number since  Majorana particles necessarily break lepton number. Further progress along these lines was cut short by his tragic disappearance.
\vskip .5cm\noindent
Starting from  Maria Goeppert-Mayer's $1935$  study of double $\beta$ decay, Wendell Furry (Phys. Rev 56, 1184(1939)) applied the Majorana idea to a similar decay neutrinoless double $\beta$ decay  ($\beta\beta^{}{{\scriptstyle 0}\nu}$) with the difference that the two electrons are expelled without their their usual antineutrinos. Furry's process gave reality to the Majorana or non Majorana nature of neutrinos. 
\vskip .5cm\noindent
In $1945$ Bruno Pontecorvo proposes a way to look for neutrinos

$$\nu_e+{}_{}^{37}Cl~~\longrightarrow~~{}^{37}_{}Ar+e^-_{}$$
Whenever a  neutrino hits a vat of cleaning fluid $CCl_4$,  an Argon isotope and an electron are produced. The beauty of the reaction is that  Argon is chemically inert and is radioactive with a half life of the order of one month with provides a beautiful  signature. Pontecorvo approached his teacher Fermi who said that although it was a nice idea, it will never be seen because the rates are so low. So it remained a preprint from Chalk River, the Canadian reactor laboratory where Pontecorvo was working. Being classified, it was not published;  even when declassified a few years later Pontecorvo did not submit it for publication\footnote{When I met Pontecorvo (once) at Erice, he gave me a reprint of his paper}.  
\vskip .5cm\noindent
 Pontecorvo's elegant reaction had not escaped Ray Davis' attention, whose skills as a radio chemist were taylor-made for this experiment. He proposes a pilot experiment near the same Savannah river nuclear plant, which generates plenty of antineutrinos but no neutrinos.
 
 A rumor soon appears according to which Davis had detected one neutrino event. Rumors propagate faster than the speed of light since they contain no information. Sure enough the rumor was just that but it had the unintended effect to motivate Pontecorvo with another  beautiful  idea (J. Exptl. Theoret. Phys. (U.S.S.R.) 33, 549 (1957)): could it be that a reactor  antineutrino  oscillates into Davis' neutrino? He reasoned by analogy with  the analysis of the neutral Kaon anti-Kaon system the year before.
 
\vskip .2cm
\hskip -1cm 
\includegraphics[scale=.3]{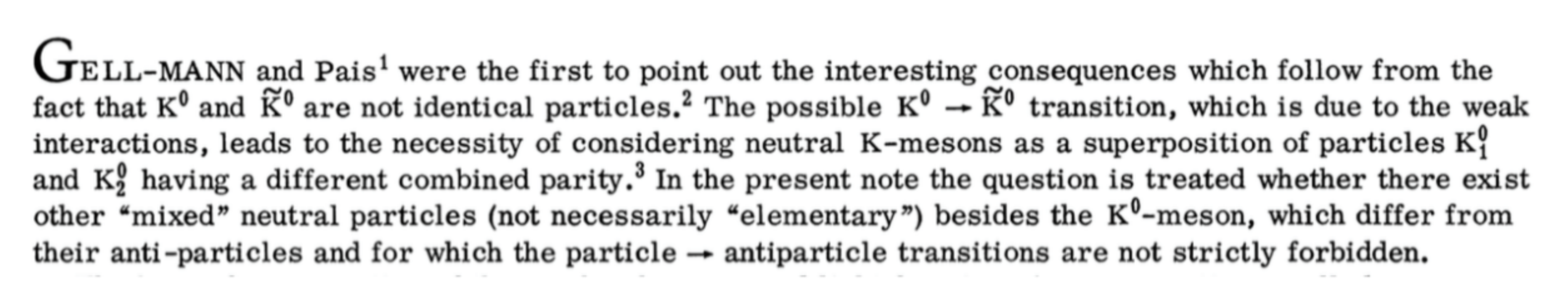}
\vskip .5cm 
\hskip -1cm 
\includegraphics[scale=.3]{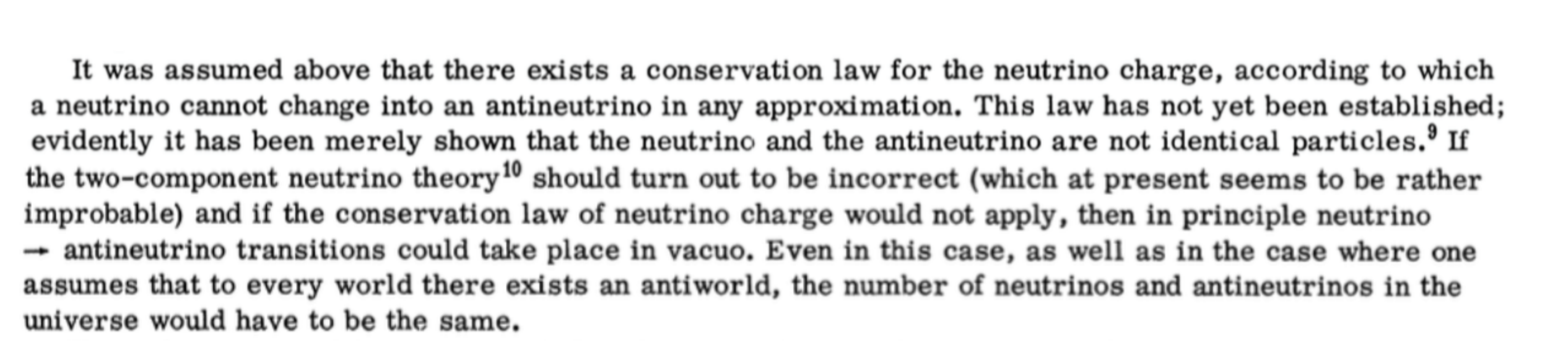}
\vskip .2cm \noindent Thus was born the idea of vacuum neutrino-antineutrino oscillations (``transmutations").

\vskip 1cm\noindent
After the Cowan-Reines experiments it was soon realized on harmonious grounds that there must be a different neutrino associated with the muon. Shoichi Sakata, Ziro Maki and Masami Nakagawa (Prog. Theo. Phys. 28, 870(1962))  applied the flavor mixing ideas of Gell-Mann and Levy to neutrinos\footnote{Kobayashi and Maskawa who discovered CP violation in quark mixing were students at Nagoya University where Sakata had extended his egalitarian ideas to particle mixings}
\vskip .5cm
\hskip .5cm 
\includegraphics[scale=.3]{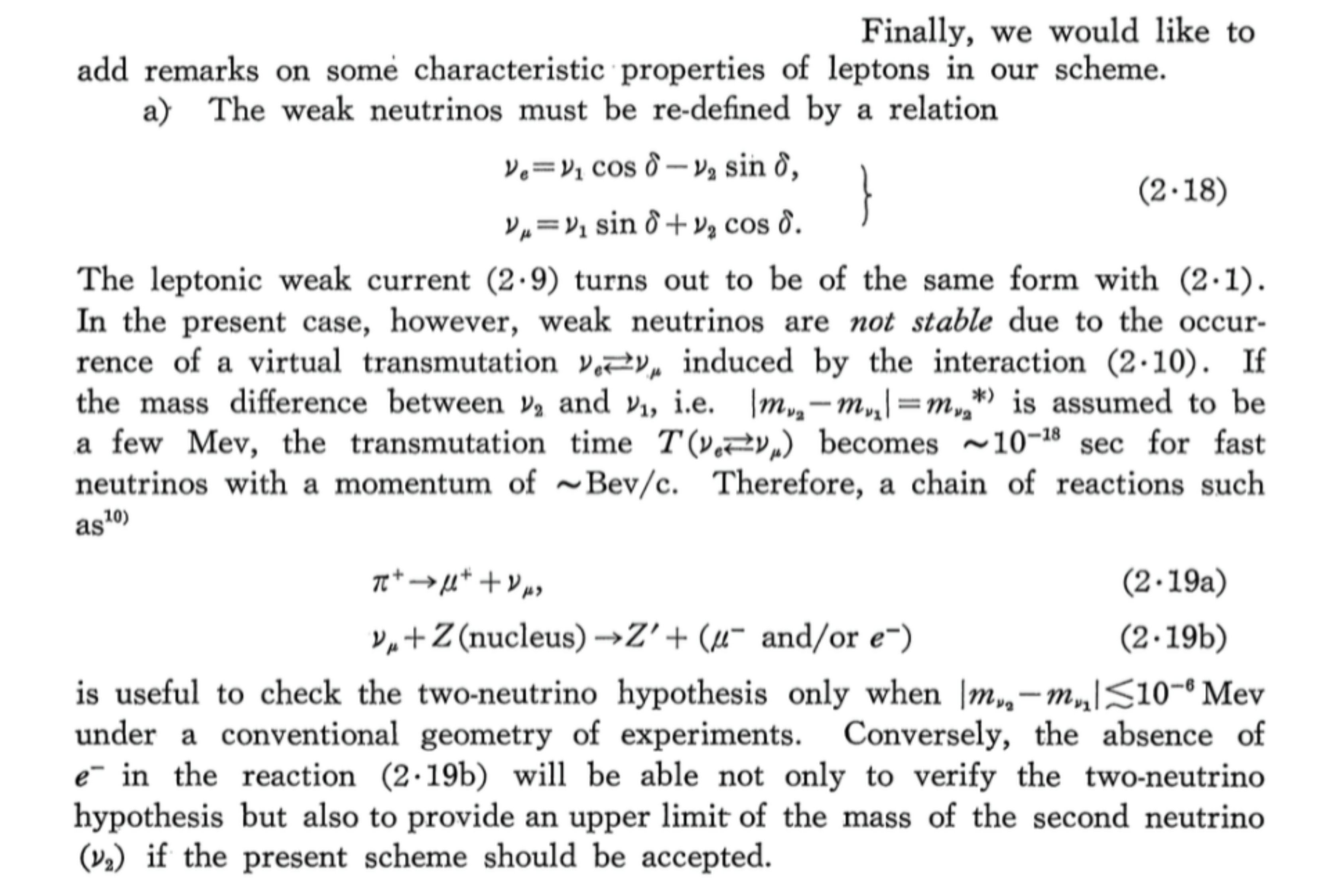}
\vskip .5cm\noindent They refer to the transmutation between the two flavors of neutrinos $\nu_e$ and $\nu_\mu$. Thus was born the idea of vacuum flavor oscillation. 
 \vskip .5cm
This concludes my short and selective description of neutrino prehistory. 

 \section{Neutrino Masses}
 It was Fermi who first attempted to determine the neutrino mass from the continuous spectrum of the $\beta$ electron. He proposed to look at the electron's spectrum at the end of its kinematically allowed range\footnote{An early example of extreme kinematics  used today to distinguish  different topologies of LHC events}. 
 
 \vskip .5cm\noindent Fermi had first published his findings in Italian ({\sl La Ricerca Scientifica}, 2, fasc. 12(1933)), and a year later in a German journal (Z.Phys. 88 161(1934)), which explains the two erroneous but suggestive figures:
 \vskip .5cm
\hskip .75cm \includegraphics[scale=.3]{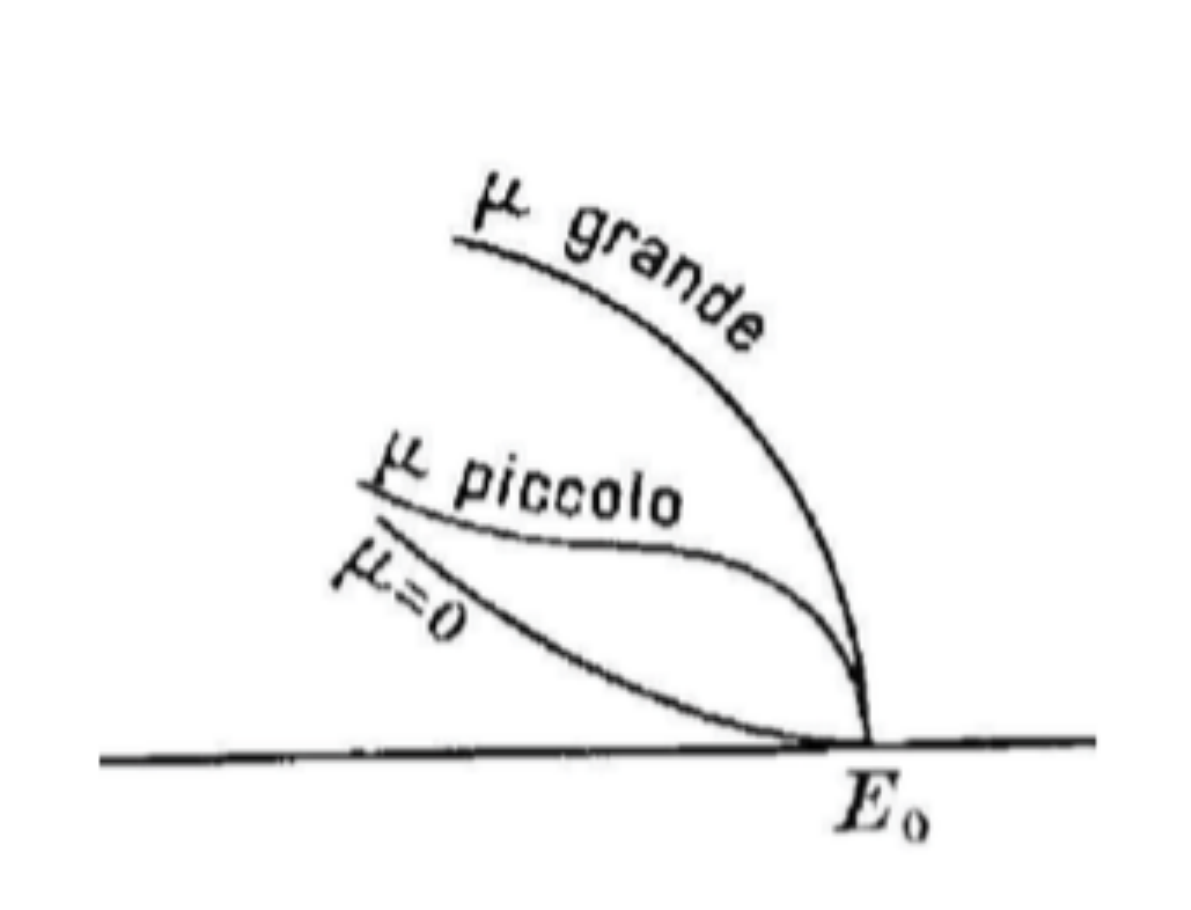} \hskip 2cm \includegraphics[scale=.3]{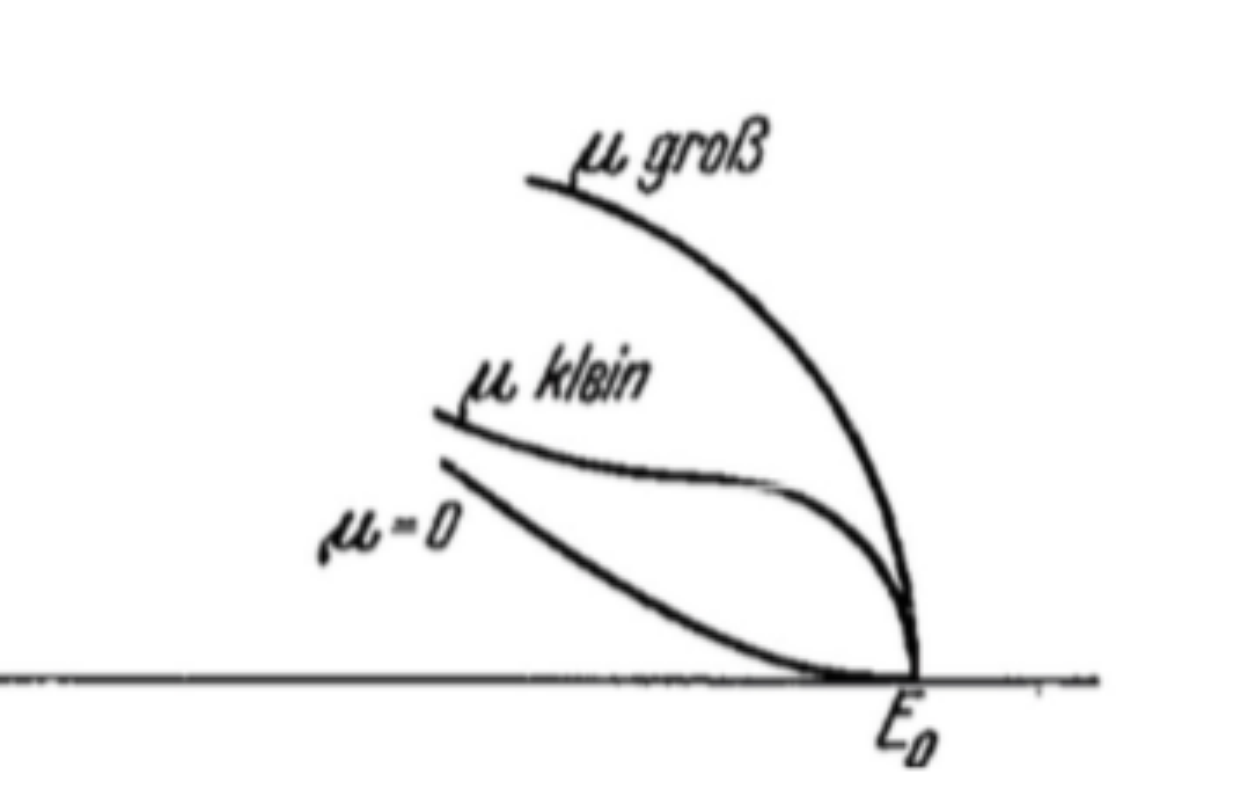}
 \vskip .5cm\noindent 
 In fact neutrinos are absurdly light, to the point that it was widely believed that they were massless\footnote{This ``what else can it be" attitude on neutrino masses is reminiscent of the cosmological constant migrating from a ``wecib" zero  to a non-zero measured value}, in which case the mixing would be irrelevant.
 \vskip 1.cm
 \noindent There are many ways to incorporate neutrino masses in the Standard Model.  All require new degrees of freedom, either bosons and/or fermions. They are distinguished by their couplings to the  three weak doublets and three weak singlets of the Standard Model leptons,
 
 $$
 L^{}_i=\begin{pmatrix}\nu^{}_i\cr e^{}_i\end{pmatrix},\qquad \bar e^{}_i,
 $$
 where $i=1,2,3=e,\mu,\tau$ is the flavor index. There are three associated global lepton numbers $\ell_i$, with  $\ell_i=+1$ for $L_i$ and $\ell_i =-1$ for $\bar e^{}_i$. 
 \vskip .5cm\noindent Neutrino masses can be generated only if new degrees of freedom, bosons and/or fermions, are added to the Standard Model.  We split the discussions into two cases  
 \vskip .5cm
\noindent  - Leptonic Bosons Only
\vskip .2cm
\noindent With no extra fermions, neutrino masses are of the Majorana type $\nu_i\nu_j\sim L_iL_j$, which break lepton numbers by two units.  Lepton-number  carrying scalars fields must be introduced.  Their renormalizable couplings to the Standard Model leptons are of three types:
\vskip .5cm  

$$\bullet~{\rm Flavor~antisymmetric}~~L^{}_{[i}L^{}_{j]}~~{\rm weak~singlets~~couple~to}~~S^+_{},$$
where $S^+$ is a charged  scalar field with hypercharge $2$ and total lepton number $\ell=\ell_e+\ell_\mu+\ell_\tau=-2$
 
$$\bullet~{\rm Flavor~symmetric}~~L^{}_{(i}L^{}_{j)}~~{\rm weak~triplets~~couple~to}~~T,$$
 where $T$ are  isotriplet scalar fields with hypercharge $2$ and total lepton number $\ell=-2$. Two of its three components  $T^{++},T^+,T^0$ carry electric charge. With two charged components, its signature makes it an experimental favorite.
 
$$\bullet~ {\rm Flavor~symmetric}~~\bar e^{}_i\bar e^{}_j~~{\rm weak~singlet~~couples~to}~~S^{--}_{},$$
where $S^{--}$ is a doubly charged scalar field. In these generic couplings, possible flavor indices are not shown.

These models break  lepton number explicitly\footnote{Spontaneous breaking generates experimentally ruled-out massless Majorons} in the potential to enable Majorana masses. In all cases explicit breaking occurs through cubic couplings of dimension three:

$$m(H\,H)\,T~~{\rm (Type~ II)},\quad mS^{--}_{}S^+_{}S^+_{},\quad mS^{--}(T\,T),\quad m\bar S^+_{}\bar S^+_{}(T\,T),
$$
and combinations thereof. All break $\ell$ by two units. The arbitrary mass parameters are determined to generate a mass suppression through mixing light and heavy states (called seesaw by some). 

There is a  model (Zee) where the neutrino masses appear at one loop. It requires a second BEH scalar $H'$ to enable the  cubic coupling $(H\,H')\,S^+_{}$, where $S^+$ couples to the flavor antisymmetric combination of two weak doublets.   

\vskip 1cm
\noindent  - Leptonic Fermions Only
\vskip .2cm
\noindent Extra fermions with lepton numbers couple renormalizably to the Standard Model in four ways using $H$ the weak doublet BEH boson (again suppressing all flavor indices):
\vskip .5cm

$$\bullet~L^{}_i\bar H~~{\rm weak~singlets~~couple~to}~~\bar N,$$
where $\bar N$ are  neutral leptons with zero hypercharge and $\ell=-1$.  Here the BEH vacuum value generates  Dirac mass terms of the form $\nu_i\bar N_j$ which does not violate total lepton number. But then why are they so small? 

$$\bullet~L^{}_i\bar H~~{\rm weak~triplets~~couple~to}~~\bar \Sigma,$$
where $\bar\Sigma$ are isotriplet fermions with zero hypercharge and $\ell=-1$. 

$$\bullet~L^{}_i H~~{\rm weak~singlets~~couple~to}~~ \bar N^+_{},$$
where $ \bar N^+$ are  charged leptons with two units of  hypercharge and $\ell=-1$. The electroweak vacuum  generates mass terms which mix $\bar N^+$ and $\bar e$. 

$$\bullet~L^{}_i H~~{\rm weak~triplets~~couple~to}~~ \vec\Sigma,$$
where $ \vec \Sigma$ are  charged leptons with two units of hypercharge and $\ell=-1$
\vskip 1cm
\noindent
 I  discuss only the first of these fermion addition models, because the extra neutral leptons can have both Dirac and Majorana masses which unite to produce a winning combination. 
 
 \noindent The $SO(10)$ Grand-Unified Theory naturally provides one fermion per chiral family, and the GUT scale generates the observed suppression  of the neutrino masses: the Seesaw Mechanism.
  
 \vskip .5cm\noindent The Dirac mass is generated in the electroweak vacuum, from $\Delta I^{}_{\rm w}=1/2$ physics at the electroweak scale $m\sim 240 GeV$. The Majorana mass with  $\Delta I^{}_{\rm w}=0$ unknown physics of unknown scale M. The three observed neutrino species have suitably suppressed masses, and the three right-handed neutrinos have masses of the order of the GUT scale. 
\vskip .5cm\noindent
The Seesaw Mechanism requires new particles with GUT scale masses: there is particle Physics Beyond the Standard Model.
\vskip 1cm

\section{Neutrino Masses and Mixings}    
The observable lepton mixing matrix results from an overlap between two types of mixings,

$$\m U^{}_{PMNS}=\m U^{\dagger}_{-1}\,\m U^{}_{Seesaw}
$$
where $\m U^{}_{-1}$  diagonalizes the charged lepton Yukawa Standard Model couplings, and 
$\m U_{Seesaw}$ diagonalizes the Seesaw matrix, of unknown $\Delta I^{}_w=0$ origin\footnote{Although any neutrino mass model could generate this matrix, I consider only the Seesaw Mechanism where the scale is motivated by Grand-Unification}.
\vskip .5cm

 \noindent Experimental neutrino mixing angles are a combination of two values,
 
 $$\theta^{}_{Expt}\sim \theta^{}_{EW}``+"\theta^{}_{Seesaw}$$
where $\theta_{EW}$ is expected to be like quark mixings,  of the order of Cabibbo angle, a sort of ``Cabibbo Haze" correction to the Seesaw mixing $\theta_{Seesaw}$. 
\vskip 1cm
The  neutrino masses are constrained by both oscillation experiments and the early universe. Oscillations data  (normal hierarchy, PDG values) yield:

\bean
\Delta^2_{12}&\equiv& \large |m^2_{\nu_1}-m^2_{\nu_2}\large |=(8.68~meV)^2,\\
\Delta^2_{23}&\equiv& \large |m^2_{\nu_1}-m^2_{\nu_3}\large |=(50.10~meV)^2.
\eean
They suggest either the ``normal hierarchy" with $m_{\nu_1}<m_{\nu_2}\ll m_{\nu_3}$, or the ``inverted hierarchy"  $m_{\nu_1}<m_{\nu_2}\gg m_{\nu_3}$ , although the former appears slighly favored. 
\vskip .5cm
\noindent The energy  in neutrino masses  in the very early universe is limited to

$$m_{\nu_1}+m_{\nu_2}+m_{\nu_3}\leq 220~meV.$$
\vskip 1cm
\noindent The measured three lepton mixing angles, 
\bean
\theta^{}_{23}&=&40.2^\circ \begin{matrix}\scriptstyle +1.4^\circ\cr \scriptstyle -1.6^\circ\end{matrix}~~\rm ``atmospheric~angle"\\
\theta^{}_{12}&=&33.6^\circ \pm .8^\circ~~\rm``solar~angle"\\
\theta^{}_{13}&=&8.37^\circ \pm .16^\circ<\theta_{Cabibbo}~~\rm``reactor~ angle"
\eean
display two large angles and a small angle less than Cabibbo's. The two large angles were unexpected while the reactor angle falls in line with naive expectations.
\vskip .5cm
\noindent The present data tends towards a $CP$-violating phase in the PMNS matrix. 
\vskip .5cm
\noindent The Seesaw mechanism predicts two other phase angles linked with Majorana physics that violate total lepton number. There is no sign of total lepton number violation in the data. 

\newpage
\section{Neutrinos \& Yukawa Unification}    
In his famous (but forgotten) lecture\footnote{Proceedings of the Royal Society (Edinburgh), Vol 59,1938-39, Part II pp. 122-129} for the James Scott Prize ``The Relation between Mathematics and Physics", Dirac discusses the principles of simplicity and mathematical beauty. Simplicity is Newton's equation while mathematical beauty is the symmetry special relativity. He even goes as far as saying 
\vskip .5cm
\noindent {\it ``It often happens that the requirements of simplicity and of beauty are the same, but where they clash the latter must take precedence"}.
\vskip .5cm
\noindent  We follow Dirac's path in  search of an organizing principle for Yukawa couplings. 
\vskip .5cm
\noindent  
Beauty can be found in the quarks and leptons gauge couplings which suggest a unifying gauge symmetry at much shorter distances.
\vskip .5cm
\noindent Neither simplicity nor beauty is easily discerned in the masses and mixings of quarks and leptons.  
\vskip .5cm\noindent Quark masses and charged leptons are strongly hierarchical; neutrino masses are not. Quark mixings are small; neutrino mixings contain two large mixings. 
\vskip .5cm
 \noindent Large angles suggest a crystal-like symmetry for a hypothetical Majorana crystal. 
 \vskip .5cm 
\cl{ Can Dirac beauty emerge from a discrete symmetry?\footnote{Discrete  flavor symmetry, advocated long ago by Sugawara and Pakvasa, and also Ma, is now hugely popular} }
 \vskip .5cm
 \noindent Three chiral families suggest  finite subgroups of $SU(3)$. These were catalogued  by mathematicians more than a century ago, and it is fair to say that each possibility can be found in the literature! 
  \vskip .5cm
 \noindent There is no compelling argument in favor of one group over another. 
 \vskip .5cm
 \noindent 
 For the remainder of this talk  I will be within this theorist's  ``Rabbit Hole" and single out the  mathematically ubiquitous  simple discrete $SU(3)$ subgroup with $168$ elements, $PSL(2,7)$.  
 \vskip .5cm
 \noindent
 It is useful to introduce a graphical rendition that shows how the different Yukawa couplings of the Standard Model are connected by Grand-Unification.
 
 \newpage

\noindent Let us represent the Standard Model  Yukawa couplings  by circles labelled by the  particles whose masses they generate. The fourth circle is the electroweak Dirac mass Yukawa with  one right-handed neutrino per chiral family.
\vskip .5cm
\cl{\includegraphics[scale=.25]{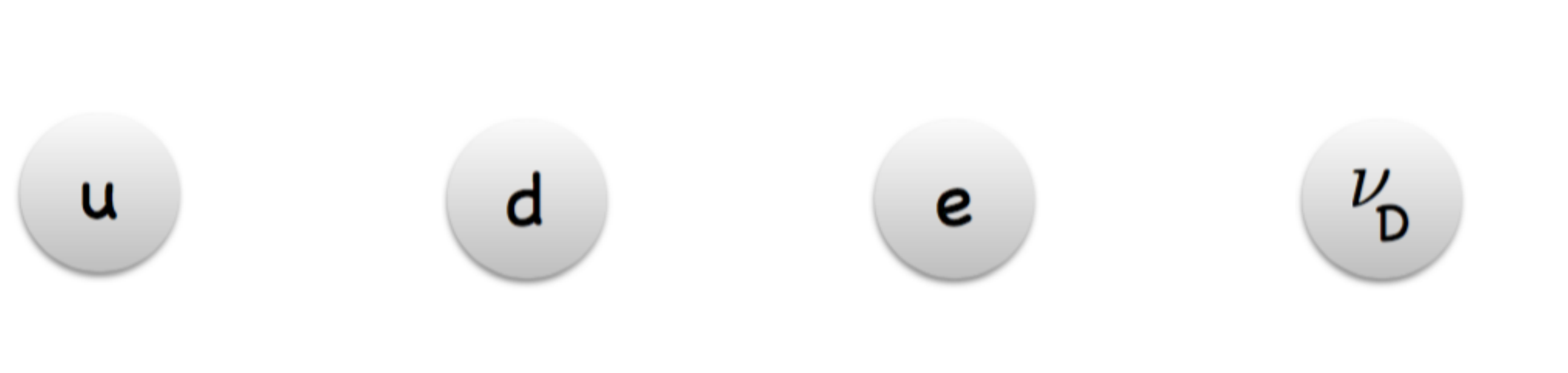}}
\vskip .5cm
\noindent
 To enable the Seesaw Mechanism we add the Majorana mass

\cl{\includegraphics[scale=.25]{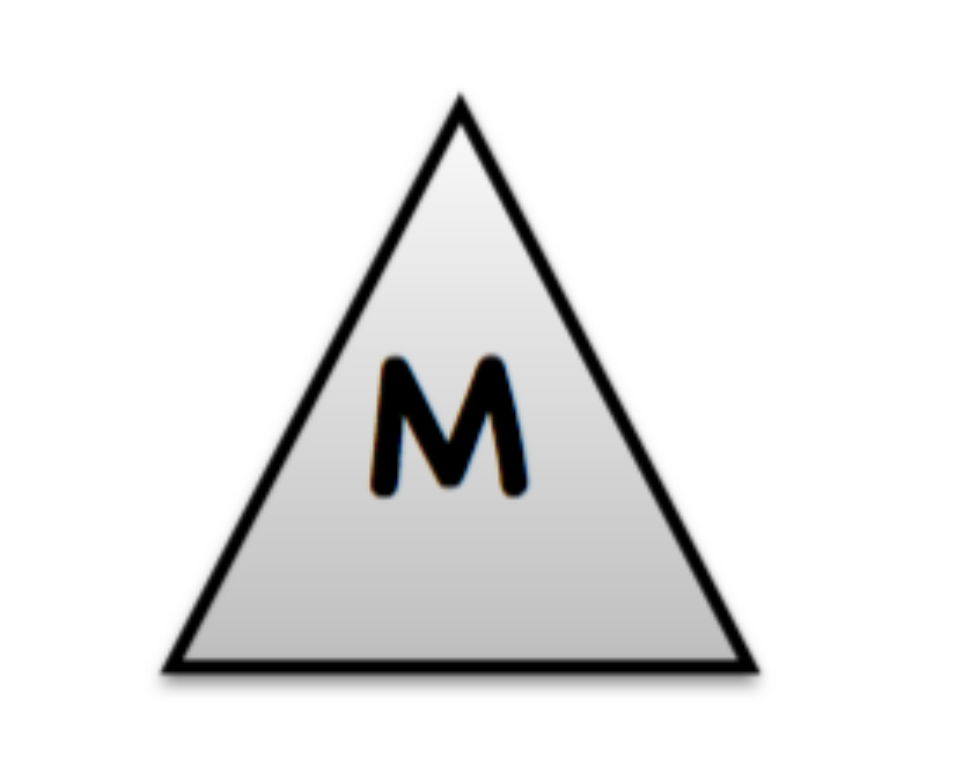}}
\vskip .5cm
\noindent The $SU(5)$ and $SO(10)$ Grand-Unified groups connect these couplings through the ``Flavor Ring" where the red links are GUT-inspired and the observable mixing matrices are the black links:

\cl{\includegraphics[scale=.3]{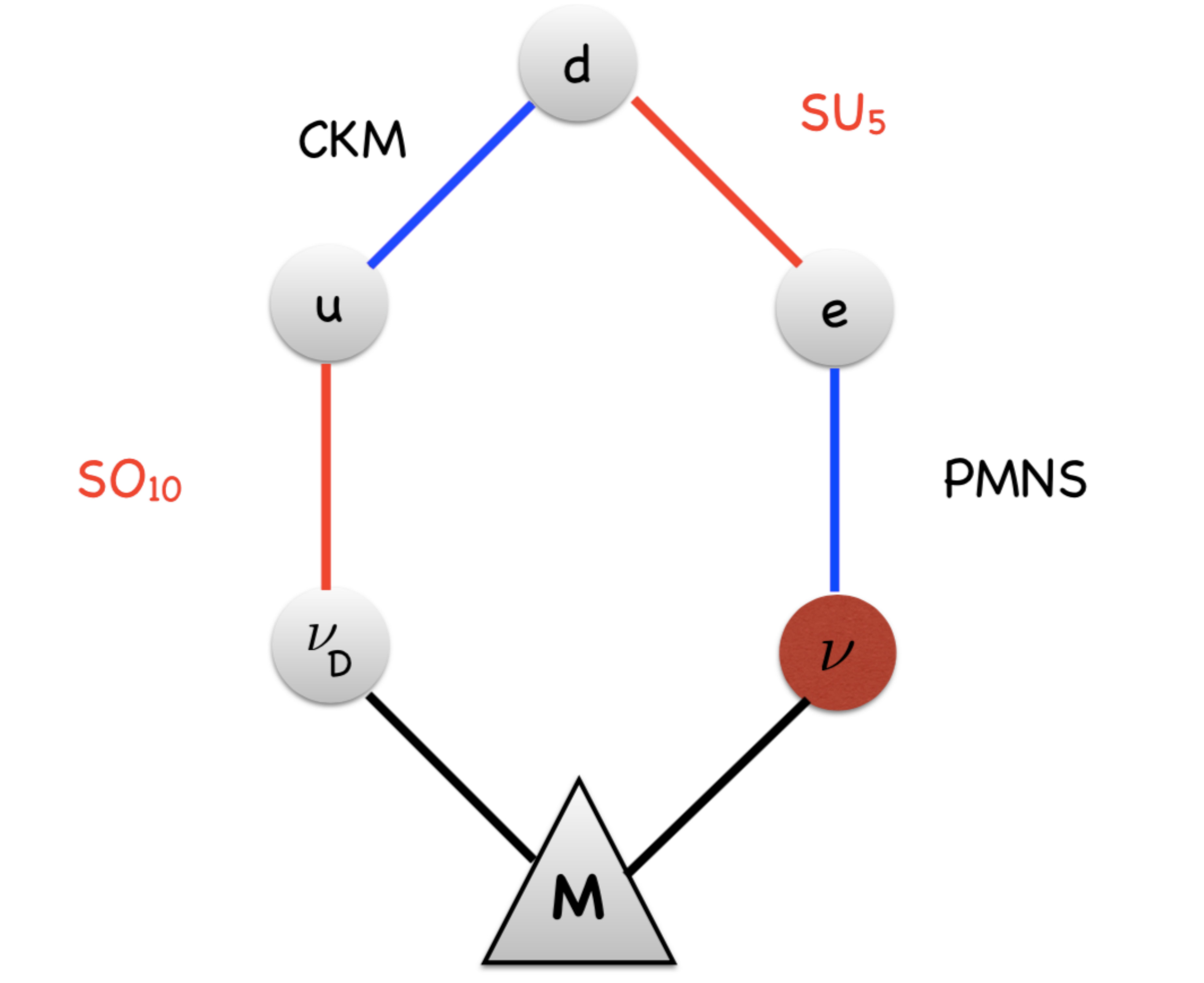}}

\newpage
\cl{}

\noindent We invoke a principle of {Seesaw Simplicity}, which posits that 
\vskip .5cm
\noindent the two large (solar and atmospheric) angles come solely from the Seesaw side,
\vskip .2cm
\noindent the small reactor angle is entirely due to the charged lepton mixing matrix.
\vskip .2cm
\noindent An obvious choice for the Seesaw mixing matrix is the ``Tri-Bi-Maximal Matrix"\footnote{``The ugly matrix with a pretty name" (L. Everett) } of Perkins et al:

$$\begin{pmatrix}\sqrt{2/3}&\sqrt{1/3}&0\cr -\sqrt{1/6}&\sqrt{1/3}&\sqrt{1/2}\cr \sqrt{1/6}&-\sqrt{1/3}&\sqrt{1/2}\end{pmatrix}
$$
\vskip .2cm
\noindent  Seesaw simplicity is most clearly enunciated when the Yukawa matrix of the charge $2/3$ quarks is diagonal: $Y^{2/3}\sim{\rm Diag}(\epsilon^4,\epsilon^2,1)$\footnote{Natural when the family symmetry distinguishes diagonal from off-diagonal  couplings}
\vskip .5cm
\noindent $\bullet~SO(10)$ link:  charge $2/3$ and  Dirac neutrino mass matrices are equal at GUT scale
\vskip .2cm
\noindent $\bullet~SU(5)$ link: $m_b=m_\tau$ determines the GUT scale $M_{\rm GUT}$ using the renormalization group\footnote{Possible only because the $b$ quark  physical mass (half the $\Upsilon$) is bigger than  the  $\tau$ lepton's}.
\vskip .5cm
\noindent Absence of dramatic hierarchy in neutrino masses $\rightarrow$  ``correlated hierarchy" in the Majorana mass matrix  

$${\m M}=\begin{pmatrix} \epsilon^4&0&0\cr 0&\epsilon^2&0\cr 0&0& 1\end{pmatrix} {\m M'} \begin{pmatrix} \epsilon^4&0&0\cr 0&\epsilon^2&0\cr 0&0& 1\end{pmatrix}$$
$\m M'$ of order one with inverse eigenvalues proportional to neutrino masses. TBM diagonalization fixes relations among its elements:

$${\m M'_{12}}={\m M'_{13}};\quad {\m M'_{22}}={\m M'_{23}};\quad{\m M'_{11}}+{\m M'_{12}}+{\m M'_{23}}={\m M'_{22}}
$$
\vskip .5cm
\noindent Choice of discrete group is predictive (G. Chen, J. M. P\'erez):

$$PSL(2,7)~~\rightarrow~~{\m M'_{22}}={\m M'_{23}}~~\rightarrow~~\Big|\frac{m_{\nu_1}}{m_{\nu_2}}\Big|=\frac{1}{2}$$
Folding this extra relation with the oscillation data yields 

$$m_{\nu_3}\sim~ 50~ meV,\quad m_{\nu_2}\sim 11~ meV,\quad m_{\nu_1}\sim 5.5~ meV$$

\newpage
\noindent The $2014$ Florida flavor group (J. Kyle, J. M. P\'erez, J. Zhang) found that TBM mixing required  flavor-asymmetric charged lepton Yukawa matrices.
\vskip .5cm\noindent 
Recently  my students (M.H. Rahat and Bin Xu) and I  presented a TBM texture 
that fits the GUT patterns and all mass and mixing angles data but only for a specific CP-violation.

\vskip .5cm \noindent 
$\bullet~SU(5)$ relate charge $-1/3$ and charge $-1$ Yukawa matrices with BEH along the $\bf \bar 5$ and $\bf\overline {45}$  representations. 

The Yukawa matrices are expressed in terms of the Wolfenstein parameters $A,\rho,\eta,\lambda$

\bean
{\bf\bar 5}&:&~~~\frac{1}{3}\begin{pmatrix} 2\sqrt{\rho^2+\eta^2}\lambda^4&\lambda^3&3A\sqrt{\rho^2+\eta^2}\lambda^3\cr
\lambda^3&0&3A\lambda^2\cr
3A\sqrt{\rho^2+\eta^2}\lambda^3&3A\lambda^2&3
\end{pmatrix}+\frac{2\lambda}{3A}\begin{pmatrix} 0&0&0\cr 0&0&0\cr 1&0&0\end{pmatrix}
\\
{\bf\overline {45}}&:&~~~\frac{\lambda^2}{3}\begin{pmatrix} 0&0&0\cr 0&1&0\cr 0&0&0\end{pmatrix}
\eean
They reproduce the Wolfenstein CKM matrix, the Gatto relation and the GUT-scale Georgi-Jarlskog relations

$$
\lambda\approx \sqrt{\frac{m_d}{m_s}},\quad m_b=m_\tau,\quad m_\mu=3m_s,\quad m_d=3m_e.$$
The PMNS angles are also determined, but they differ from their PDG values,

$$\theta_{13}:~ 2.26^\circ {\rm above~pdg},\quad \theta_{23}:~ 2.9^\circ {\rm below},\quad\theta_{12}:~ 6.16^\circ {\rm above}.
$$
These angles  can be brought back to their PDG  values by adding a CP-violating  phase $\varphi$ in the TBM matrix. This is possible because the reactor angle is above its experimental value. 

Lowering  the reactor angle to its PDG value demands $\cos\varphi\approx 0.2$, but leaves the sign of $\varphi$ undetermined. The other two angles magically fall within PDG:

$$\cos\varphi\approx 0.2~\rightarrow~\theta_{13}~{\rm at ~pdg},\quad \theta_{23}:~ 0.66^\circ {\rm below},\quad\theta_{12}:~ 0.51^\circ {\rm above}.
$$
 The Jarlskog-Greenberg invariant is $J=|0.027|$. When folded into the PMNS matrix, we find  
$\delta^{}_{CP}=1.32\pi$ or $\delta^{}_{CP}=0.67\pi$, depending on the sign of the phase. Only the first value is consistent with the little we know from experiments.

\newpage
\centerline {A Neutrino Prediction}
\vskip 1cm
\noindent Two important measurements await neutrino physics, neutrinoless double $\beta$-decay which will determine if the total lepton number is broken, and the cosmic neutrino background. In the absence of technology which suggests these measurements in the near future, I turn whimsically to mathematics to offer a prediction for the year when lepton number violation will be detected:

\bean
  {\rm Revelation}&:&  \hskip 3.67cm 1930\\
{\rm Detection}&:& ~26=2\cdot13~{\rm years~ later}~~ 1956\\
{\rm Oscillations}&:&  ~68=2^2\cdot 17~{\rm years~ later}~ 1998\\ 
\beta\beta_{{\scriptstyle 0}\nu}~ {\rm Decay}&:&  152=2^3\cdot 19~{\rm years~ later}~ 2052
\eean
\vskip 2cm

\centerline{\Large The Sun Never Sets }
\vskip .5cm
\hskip 1.5cm \includegraphics[scale=.3]{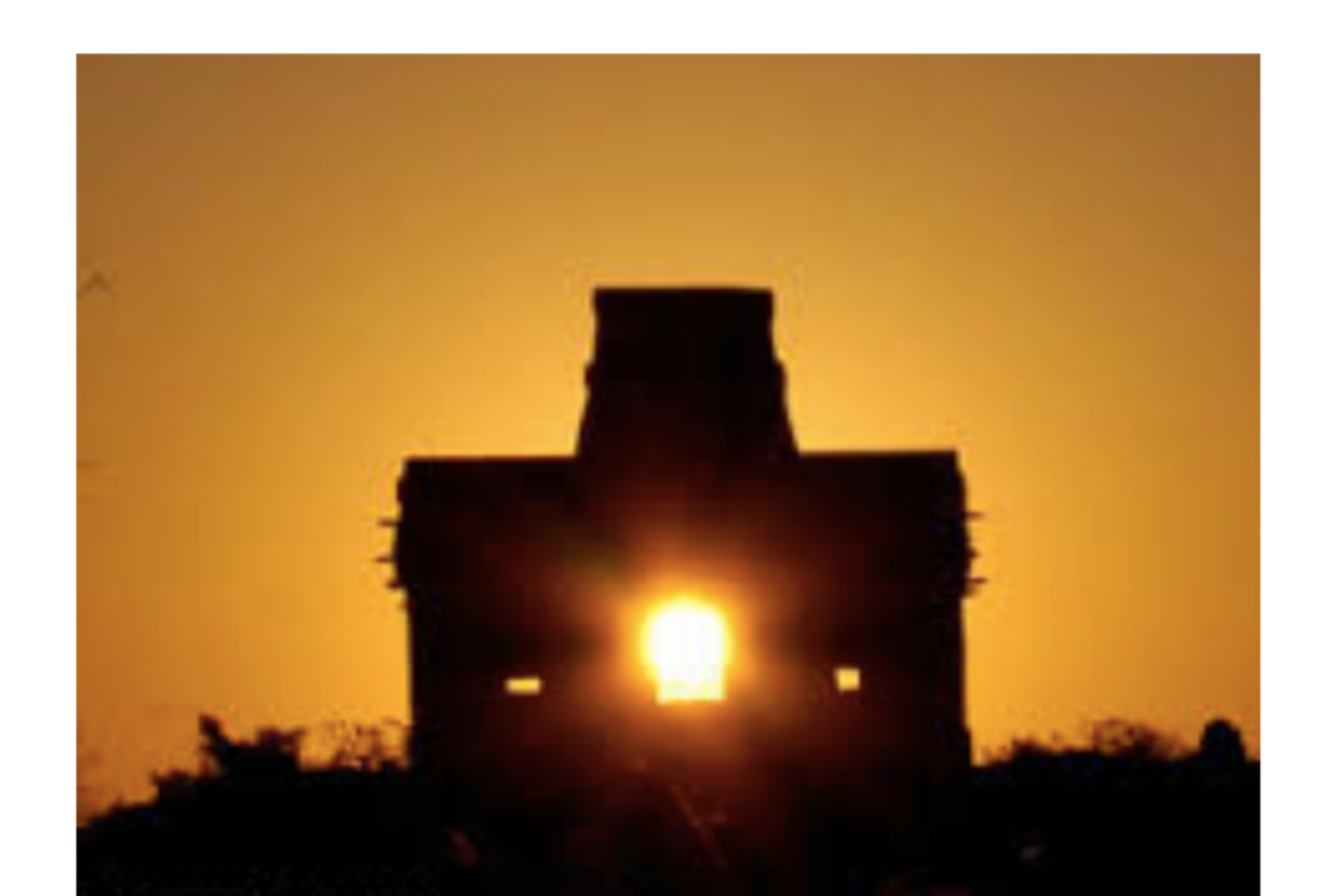}
\vskip .5cm
\centerline{\Large On Neutrino Detectors}
\vskip1cm
\noindent I thank Professors Daniel Vignaud and Michel Cribier for their kind invitation, the Aspen Center for Physics for its hospitality, and the support of the Department of Energy under Grant No. DE- SC0010296. 
\end{document}